\documentclass[11pt,a4paper]{amsart}
\usepackage[utf8]{inputenc}
\usepackage[T1]{fontenc}
\usepackage{lmodern}
\usepackage[margin=28mm]{geometry}
\usepackage{amsmath,amssymb,amsthm,mathrsfs}
\usepackage{booktabs,array,microtype}
\usepackage{xcolor}
\usepackage{cite}
\usepackage{enumitem}
\usepackage{graphicx}
\usepackage{tikz}
\definecolor{cT}{HTML}{f39c34}\definecolor{cD}{HTML}{3a62d8}\definecolor{cO}{HTML}{e0261f}
\definecolor{cCw}{HTML}{2a8c82}\definecolor{cCb}{HTML}{8b3fb0}
\newcommand{\fsq}[4]{% x y fill label
  \fill[#3!35] (#1,#2) rectangle ++(1.1,1.1);
  \draw[line width=0.35pt] (#1,#2) rectangle ++(1.1,1.1);
  \draw[#3,line width=0.65pt] (#1,#2) -- ++(1.1,1.1);
  \fill[black] (#1,#2) circle (1.6pt); \fill[black] (#1+1.1,#2+1.1) circle (1.6pt);
  \filldraw[fill=black!35,draw=black] (#1+1.1,#2) circle (1.6pt); \filldraw[fill=black!35,draw=black] (#1,#2+1.1) circle (1.6pt);
  \node at (#1+0.25,#2+1.1-0.22) {#4};
  \node[left] at (#1,#2) {\tiny$0$}; \node[right] at (#1+1.1,#2+1.1) {\tiny$\pi/2$};}

\usetikzlibrary{patterns}
\definecolor{ck0}{RGB}{89,89,89}
\definecolor{ck1}{RGB}{230,90,30}
\definecolor{ck2}{RGB}{40,90,200}
\definecolor{ck3}{RGB}{20,120,90}
\usepackage{flafter}
\usepackage{placeins}
\usepackage[hidelinks]{hyperref}
\newcommand{\R}{\mathbb R}
\newcommand{\C}{\mathbb C}
\newcommand{\U}{\mathrm U}

\DeclareMathOperator{\diag}{diag}
\DeclareMathOperator{\tr}{tr}
\DeclareMathOperator{\Ran}{Ran}
\DeclareMathOperator{\supp}{supp}
\newcommand{\shid}[4]{{\scriptsize$\left[\begin{array}{@{}c@{\,}c|c@{\,}c@{}}#1\\#2\\\hline #3\\#4\end{array}\right]$}}
\newtheorem{theorem}{Theorem}[section]
\newtheorem{proposition}[theorem]{Proposition}
\newtheorem{lemma}[theorem]{Lemma}
\newtheorem{corollary}[theorem]{Corollary}
\theoremstyle{definition}
\newtheorem{definition}[theorem]{Definition}
\newtheorem{example}[theorem]{Example}
\theoremstyle{remark}
\newtheorem{remark}[theorem]{Remark}
\numberwithin{equation}{section}
\allowdisplaybreaks[1]
\title[Reconstruction and geometry of quantum Sudoku]
{Reconstruction and geometry of $4\times4$ quantum Sudoku}
\author[F. D. Cunden]{Fabio Deelan Cunden}
\address{Dipartimento di Matematica, Universit\`a di Bari,  I-70125 Bari, Italy and INFN, Sezione di Bari,
I-70126 Bari, Italy}
\email{fabio.cunden@uniba.it}
\author[P. Facchi]{Paolo Facchi}
\address{Dipartimento di Fisica, Universit\`a di Bari, I-70126 Bari, Italy and INFN, Sezione di Bari,
I-70126 Bari, Italy}
\email{paolo.facchi@ba.infn.it}
\author[R. La Scala]{Roberto La Scala}
\address{Dipartimento di Fisica, Universit\`a di Bari, I-70126 Bari, Italy}
\email{roberto.lascala@uniba.it}
\author[M. Ligab\`o]{Marilena Ligab\`o}
\address{Dipartimento di Matematica, Universit\`a di Bari,  I-70125 Bari, Italy}
\email{marilena.ligabo@uniba.it}
\author[K. \.Zyczkowski]{Karol \.Zyczkowski}
\address{Institute of Theoretical Physics, Jagiellonian University, Krak\'ow, Poland}
\email{karol.zyczkowski@uj.edu.pl}
\keywords{Quantum Sudoku, quantum Latin squares, fractional Latin squares,
rank-one projections, matrix completion, unitary equivalence}
\subjclass[2020]{Primary 81P45; Secondary 05B15, 15A83, 15B51, 52B12}

\begin{document}
\raggedbottom

\date{November 10, 2026}

\begin{abstract}
We study the reconstruction and geometry of $4 \times 4$ quantum Sudoku, viewed as arrays of rank-one projections whose rows, columns and blocks decompose the identity. We give necessary and sufficient conditions for two prescribed opposite blocks to admit a completion, prove its uniqueness, and obtain explicit reconstruction formulas, also expressed in the language of quantum logic. The sixteen overlaps between these blocks determine the unitary equivalence class of the grid, while the number of nonzero overlaps equals its cardinality. Using a matrix representation, we parametrise the equivalence classes by five closed squares, with the sides of one square identified with the diagonals of the other four. The quotient by classical cell symmetries is a closed disc containing exactly two classical classes. Finally, measurement in the basis of the first block gives a fractional Sudoku that retains the full unitary class and is a convex combination of at most four classical grids. This measurement map embeds the five-square model into a seven-dimensional polytope whose vertices are the twelve classical Sudoku.
\end{abstract}
\maketitle

\section{Introduction}\label{sec:introduction}
A Latin square arranges $n$ symbols in an $n\times n$ grid so that each
symbol occurs once in every row and column. Replacing the symbols by
the canonical basis vectors $e_1,\ldots,e_n$ of $\C^n$ expresses this
rule as orthonormality: every row and every column is an orthonormal
basis. A \emph{quantum Latin square} retains this condition but allows
the basis to vary from one row or column to another~\cite{MV}.
The entries are unit vectors, and vectors differing only by a phase
represent the same state.

For example, let $e_1,\ldots,e_4$ be the canonical basis of $\C^4$ and put
\[
u_+=\frac{e_1+e_2}{\sqrt2},\qquad
u_-=\frac{e_1-e_2}{\sqrt2}.
\]
The grid
\begin{equation}\label{QLS}
\begingroup
\renewcommand{\arraystretch}{1.3}
\setlength{\arraycolsep}{10pt}
\left[\begin{array}{cccc}
e_1&e_2&e_3&e_4\\
e_2&e_1&e_4&e_3\\
e_3&e_4&u_+&u_-\\
e_4&e_3&u_-&u_+
\end{array}\right]\endgroup
\end{equation}
is a quantum Latin square: each row and column contains either the
canonical basis or the orthonormal basis $e_3,e_4,u_+,u_-$. It is
nonclassical even up to a common unitary change of basis. Indeed,
$|\langle e_1,u_+\rangle|^2=1/2$, whereas two states drawn from a single
orthonormal basis have squared overlap either zero or one. This
illustrates how orthogonality constraints can coexist with states
that are neither identical nor orthogonal when taken from different
rows and columns. Quantum Latin squares connect combinatorial designs
with unitary error bases and mutually unbiased bases~\cite{MV,MustoMUB,MVOrth,Zang_2021}. Entanglement also allows a quantum solution of Euler’s problem of thirty-six officers, which has no classical solution~\cite{RBB2022}.

A quantum Sudoku, or \emph{SudoQ}, adds the block constraint of the
classical puzzle~\cite{NP}. A grid has $N^2\times N^2$ cells, divided
into $N^2$ blocks of $N\times N$ cells; the vectors in each block
must also form an orthonormal basis of $\C^{N^2}$. For convenience,
we call such a grid a SudoQ \emph{of size $N$}. In particular, the
$4\times4$ SudoQ studied here have $2\times2$ blocks and are called
\emph{SudoQ of size two} throughout the paper.
The quantum Latin square in (\ref{QLS}) does not satisfy this
additional condition: its upper-left $2\times2$ block repeats $e_1$
and $e_2$. With the same six states, however, the arrangement
\begin{equation}\label{eq:intro-sudoq-six}
\begingroup
\renewcommand{\arraystretch}{1.3}
\setlength{\arraycolsep}{10pt}
\left[\begin{array}{cc|cc}
e_1&e_2&e_4&e_3\\
e_3&e_4&u_-&u_+\\\hline
e_4&e_3&u_+&u_-\\
e_2&e_1&e_3&e_4
\end{array}\right]\endgroup
\end{equation}
satisfies all three requirements. Every row, column and block is an
orthonormal basis, and the grid contains exactly six distinct states
up to phases. Its \emph{cardinality} is therefore six, rather than the
four of a classical Sudoku of this size. More generally, replacing
$u_+,u_-$ by the orthonormal pair
\[
u_+(\theta)=\cos(\theta)\,e_1+\sin(\theta)\,e_2,\qquad
u_-(\theta)=\sin(\theta)\,e_1-\cos(\theta)\,e_2
\]
preserves the Sudoku conditions. The cardinality is six for
$0<\theta<\pi/2$.
Paczos et al.~\cite{KZ} parametrised all the $4 \times 4$ quantum Sudoku and established that their possible
cardinalities are $4,6,8,16$. Constructions of mutually orthogonal
quantum Sudoku squares in higher dimensions were subsequently given
in~\cite{ZZTF}. 

In this article we study how the solutions are
reconstructed from partial data and how their unitary equivalence
classes fit together. Which data determine a grid? How is its
cardinality encoded in those data? What is the geometry of the
resulting space, including the classical grids and the transitions
between different cardinalities?

To address these questions we use the projection formulation of SudoQ~\cite{NP}: a unit vector
$u$ in a cell is replaced by the rank-one projection $uu^*$. This
removes its arbitrary phase without changing the orthogonality
conditions or the cardinality. A collection of $N^2$ unit vectors
forms an orthonormal basis of $\C^{N^2}$ precisely when the
corresponding rank-one projections sum to the identity. Thus the
Sudoku conditions say that each row, column and block is an
orthogonal decomposition of the identity into rank-one projections.

Two grids of unit vectors are unitarily equivalent when the same unitary operator
$\Gamma$ sends the state in every cell of one grid to the state in
the corresponding cell of the other, up to an individual phase.
In projection notation this means
$P'_{r,c}=\Gamma P_{r,c}\Gamma^*$ for every cell $(r,c)$.
It is a single change of orthonormal basis for the whole grid.

For SudoQ of size two, we show that the sixteen overlaps between
two opposite blocks form a complete invariant of unitary equivalence.
Taking the first and fourth blocks, these are
\[
\tr(P_{a,b}P_{2+r,2+c}),\qquad a,b,r,c\in\{1,2\}.
\]
Thus two grids are unitarily equivalent precisely when their
corresponding overlaps agree. The same data also determine the
cardinality: it is exactly the number of nonzero overlaps.

We also address the reconstruction of a SudoQ from two prescribed opposite blocks. Suppose that the projections in
the upper-left and lower-right blocks are prescribed, each block
being an orthogonal decomposition of the identity. Can the eight
remaining cells be filled so that every row, column and block also
sums to the identity? The conditions within the two prescribed
blocks alone do not guarantee a completion: the states in those
blocks must also satisfy compatibility conditions imposed by the
rows and columns. We give necessary and sufficient conditions for
such a completion and prove that, whenever it exists, it is unique.
Each missing state must satisfy the orthogonality constraints imposed
by its prescribed row and column entries. Our criterion ensures that
the corresponding allowed subspaces intersect in a one-dimensional
space and that the resulting projections satisfy all the Sudoku
conditions, yielding explicit reconstruction formulas. This also
gives a formulation in terms of quantum logic: projections represent
propositions, and the intersection expresses the logical conjunction
of the row and column requirements.

To describe the equivalence classes explicitly, we pass from
projections to matrices. 
To obtain a matrix representation of a SudoQ of size two, we choose a unit vector in the range of each projection and arrange the four vectors of each block as the columns of a matrix, following the cell order from left to right and from top to bottom. Since the projections in each block form an orthogonal decomposition of the identity, these columns form an orthonormal basis, so the resulting matrices are unitary. A common unitary change of basis makes
the first block canonical, yielding the representation 
\[
\begin{array}{c|c}
I_4&U\\\hline
V&W
\end{array}.
\]
We prove that the unitary equivalence class is uniquely determined by the class of its fourth-block matrix $W$ under the following relation on
$\U(4)$:
\[
W\sim W'\quad\Longleftrightarrow\quad
W'=D_LWD_R\quad\text{for diagonal unitaries }D_L,D_R.
\]
The right factor changes the phases of the column vectors, while
the left factor is a common unitary transformation that preserves
the canonical projections in the first block. Two normalised SudoQ
are unitarily equivalent precisely when their fourth-block matrices
are equivalent under this relation. Not every class in
$\U(4)/{\sim}$ admits a Sudoku completion: the admissible classes are
characterised by the vanishing of specified $2\times2$ minors of
$W$. They form a space $\mathcal W$ in bijection with the unitary
equivalence classes of SudoQ of size two.

We give explicit coordinates on $\mathcal W$ by solving these matrix
conditions modulo $\sim$. Every
class has a real representative in one of five families, described
by two angles $\alpha,\beta\in[0,\pi/2]$. Within each family the
angles are unique, so its parameter domain is a closed square.
Different families can represent the same class at particular
parameter values: each side of one square coincides with a diagonal
of one of the other four. Identifying these coincident classes gives
the five-square model of the space of unitary equivalence classes. The twelve distinct corners
are the classical grids, while the interiors and edges describe
the nonclassical ones. A further part of the classification
concerns the symmetries inherited from classical Sudoku~\cite{AFLT}:
exchanging rows within a band of blocks, exchanging columns within
a stack of blocks, permuting entire bands or stacks, and transposing
the grid. These operations preserve the row, column and block
constraints and act on the space $\mathcal W$ of unitary equivalence
classes. Symbol relabellings of classical grids are already included
in unitary equivalence. We determine the quotient by these positional
symmetries and show that the five-square model becomes a closed disc.
Its two classical points correspond to the two essentially distinct
classical Sudoku of size two~\cite{AFLT}.

We finally investigate the information about a SudoQ that can be recovered from measurement probabilities. Measuring each state in a chosen orthonormal basis produces a probability distribution in every cell. 
The Sudoku
conditions imply that each outcome has total probability one in
every row, column and block. The resulting array is a
\emph{fractional Sudoku}, a probabilistic relaxation of the classical
puzzle~\cite{DHW2000,DN2024,HZ2026}. We study this measurement map
when the measurement basis is given by the first block of the grid. We prove that these outcome probabilities determine the
unitary equivalence class of the original grid, and that the array
is a convex combination of at most four classical Sudoku. Thus a
single choice of measurement basis retains all the information
needed for the classification. We also describe the ambient convex
set: fractional Sudoku with a fixed canonical first block form a
seven-dimensional polytope whose vertices are the twelve classical
grids. The measurement map embeds the five-square model in this
polytope, making explicit which probability arrays arise from SudoQ
and how they sit among the more general fractional solutions.

\subsection{Outline of the paper} The paper is organised as follows. Section~\ref{sec:preliminaries} sets out the projection and probability formulations.
Section 3 addresses existence and uniqueness of completion from
opposite blocks, including the quantum-logical formulation.
Section~\ref{sec:cardinality} establishes how the overlaps encode both cardinality
and unitary equivalence, and gives examples illustrating all
possible cardinalities.
Section~\ref{sec:matrix-representation} passes to normalised block
matrices and angular coordinates, constructing the five-square
model. Section~\ref{sec:cell-symmetries} computes the action of
classical cell symmetries on these coordinates and derives the disc
quotient. Section~\ref{sec:realparam} develops the measurement
representation and places its image inside the polytope of fractional
Sudoku. The proofs are collected in Section~\ref{sec:proofs}.
Section~\ref{sec:conclusions} contains the conclusions and open problems. Appendix~\ref{app:example} gives a worked example: one SudoQ as a grid of
vectors, as the matrices $I_4,U,V,W$, and as a visual encoding. 

\section{Quantum Sudoku}\label{sec:preliminaries}
\emph{Notation.} $I_n$ is the identity matrix of order $n$, $\U(n)$
the unitary group, $[X,Y]=XY-YX$ the commutator, $M^{\mathsf T}$ the
transpose of $M$ and $\#S$ the number of elements of a finite set $S$.
For an orthogonal projection $Q$ we write $Q^\perp=I_n-Q$ and $\Ran Q$
for its range. All projections are orthogonal; a rank-one projection
is $uu^*$ ($|u\rangle\langle u|$ in Dirac notation) with $\|u\|=1$, and $u$ is unique up to a phase.

Fix $N\geq1$. Rows and columns are indexed by $1,\ldots,N^2$, and the $N^2$
\emph{blocks} are the sets of $N^2$ cells
\[
\{(N(i-1)+k,N(j-1)+l):1\leq k,l\leq N\},
%\qquad 1\leq i,j\leq N.
\]
indexed by $(i,j)$ with $1\leq i,j\leq N$.

\begin{definition}\label{def:SudoQ}
A \emph{SudoQ} of size $N$ is an array
$\mathbf P=(P_{r,c})_{r,c=1}^{N^2}$ of rank-one projections on
$\C^{N^2}$ such that the projections in every row $r$, column $c$ and block $(i,j)$
sum to the identity:
\begin{align*}
\sum_{c=1}^{N^2}P_{r,c}&=I_{N^2},&
\sum_{r=1}^{N^2}P_{r,c}&=I_{N^2},&
\sum_{k,l=1}^{N}P_{N(i-1)+k,N(j-1)+l}&=I_{N^2}.
\end{align*}
Its \emph{cardinality} is $c(\mathbf P)=\#\{P_{r,c}:1\leq r,c\leq N^2\}$.
\end{definition}

Equivalently, the projections are pairwise orthogonal within every row,
column and block. 

A \emph{classical Sudoku} is an $N^2\times N^2$ array $(S_{r,c})_{r,c=1}^{N^2}$ with entries in
$\{1,\ldots,N^2\}$ in which every row, column and block contains each
symbol once; replacing each symbol $a$ by $E_a=e_ae_a^*$, where
$\{e_a\}_{a=1}^{N^2}$ is the canonical basis of $\C^{N^2}$, 
namely $P_{r,c}=E_{S_{r,c}}$,
yields a \emph{classical SudoQ}. For results on classical Sudoku, see~\cite{AFLT,Felgenhauer06,McGuire14,Russell06} and the references therein.

\begin{definition}\label{defequiv}
Two SudoQ $\mathbf P$, $\mathbf P'$ of size $N$ are \emph{unitarily
equivalent}, $\mathbf P\equiv\mathbf P'$, if there is $\Gamma\in\U(N^2)$
with $P'_{r,c}=\Gamma P_{r,c}\Gamma^*$ for all $r,c \in \{1,\dots, N^2\}$.
\end{definition}

The following characterisation is proved in~\cite{KZ}; we state it
in the language of projections.

\begin{theorem}[Minimal cardinality]\label{prop:commuting-classical}
For a SudoQ $\mathbf P$ of size $N$ the following are equivalent:
\begin{enumerate}
\item $c(\mathbf P)=N^2$;
\item $\mathbf P$ is unitarily equivalent to a classical SudoQ;
\item all the projections of $\mathbf P$ commute pairwise.
\end{enumerate}
\end{theorem}

\subsection{Fractional Sudoku}\label{sec:fractional}
Measuring the states of a SudoQ produces a classical object of the
following kind.

\begin{definition}\label{def:fractional-sudoku}
A \emph{fractional Sudoku} of size $N$ is an array
$\mathbf p=(p_{r,c})_{r,c=1}^{N^2}$ of probability vectors
$p_{r,c}=(p_{r,c}(a))_{a=1}^{N^2}$, satisfying
\[
\sum_{c=1}^{N^2}p_{r,c}(a)=1,\qquad
\sum_{r=1}^{N^2}p_{r,c}(a)=1,\qquad
\sum_{k,l=1}^{N}p_{N(i-1)+k,N(j-1)+l}(a)=1
\]
for every symbol $a\in\{1,\ldots,N^2\}$, every row $r$,
every column $c$, and every block $(i,j)$, with
$i,j\in\{1,\ldots,N\}$.
\end{definition}
Thus $\mathbf p=(p_{r,c})_{r,c=1}^{N^2}$ is a fractional Latin square of order
$N^2$~\cite{DHW2000,HZ2026} whose blocks obey the same condition as
its rows and columns; this is the linear relaxation of Sudoku studied
in~\cite{DN2024}. Moreover, the entries \(p_{r,c}(a)\) form a tristochastic tensor: summing over any one of the three indices gives one for every fixed pair of the other indices. Such tensors also appear in the study of local thermal operations and classical communication~\cite{Bistron26}. A classical Sudoku $S$ is identified with the
fractional Sudoku $p_{r,c}=e_{S_{r,c}}$, where $\{e_1, \dots , e_{N^2}\}$ is the
canonical basis of $\R^{N^2}$.

If $\mathbf P$ is a SudoQ of size $N$ and $\{f_a\}_{a=1}^{N^2}$ is an
orthonormal basis of $\C^{N^2}$, then
\begin{equation}\label{eq:image}
p_{r,c}(a)=\langle f_a,P_{r,c}f_a\rangle,\qquad r,c,a=1,\ldots,N^2,
\end{equation}
is a fractional Sudoku, as one sees by applying
$\langle f_a,\,\cdot\,f_a\rangle$ to the identities of
Definition~\ref{def:SudoQ}: $p_{r,c}$ is the distribution of outcomes
when the state in cell $(r,c)$ is measured in the basis $\{f_a\}_{a=1}^{N^2}$.
These probabilities depend only on the squared moduli of the
components of a unit vector representing the state in this basis.
The result depends on the basis: for $N=2$ and
the basis $\{(H\otimes H)e_a : a \in \{1,2,3,4\}\}$, with
$H=\frac1{\sqrt2}\left(\begin{smallmatrix}1&1\\1&-1\end{smallmatrix}\right)$,
every classical SudoQ gives $p_{r,c}(a)=\frac14$, for all $r,c, a \in \{1,2,3,4\}$, which retains nothing
of the grid. Not every fractional Sudoku arises in this way, whatever
the basis.

\begin{lemma}\label{lem:unistochastic}
Let $\mathbf p$ be given by~\eqref{eq:image} for a SudoQ $\mathbf P$
of size $N$. For every row $r$ the matrix $(p_{r,c}(a))_{c,a=1}^{N^2}$ is
unistochastic, and similarly for every column and every block. In
particular, if $(r,c)\neq(r',c')$ lie in the same row, column or block,
there is never exactly one symbol $a$ with $p_{r,c}(a)\,p_{r',c'}(a)>0$.
\end{lemma}

The second statement is the elementary fact that two orthogonal
vectors cannot have exactly one coordinate in which both are nonzero.

\section{Reconstruction from two opposite blocks}\label{sect:mainth}
From now on $N=2$. Suppose that only the first and the fourth block are prescribed,
each consisting of four mutually orthogonal rank-one projections
on $\C^4$ whose sum is $I_4$:
\begin{equation}\label{eq:partial-grid}
\left[
\begin{array}{cc|cc}
P_{1,1}&P_{1,2}&?&?\\
P_{2,1}&P_{2,2}&?&?\\\hline
?&?&P_{3,3}&P_{3,4}\\
?&?&P_{4,3}&P_{4,4}
\end{array}\right].
\end{equation}

\begin{theorem}[Reconstruction from two opposite blocks]\label{thm:WUV}
The partial grid~\eqref{eq:partial-grid} admits a SudoQ completion if
and only if
\[
(P_{1,1}+P_{1,2})(P_{3,3}+P_{4,3})
\quad\text{and}\quad
(P_{1,1}+P_{2,1})(P_{3,3}+P_{3,4})
\]
are rank-one projections, that is,
\begin{align}
\bigl[P_{1,1}+P_{1,2},\,P_{3,3}+P_{4,3}\bigr]&=0,&
\tr\bigl((P_{1,1}+P_{1,2})(P_{3,3}+P_{4,3})\bigr)&=1,
\label{eq:projector-criterion}\\
\bigl[P_{1,1}+P_{2,1},\,P_{3,3}+P_{3,4}\bigr]&=0,&
\tr\bigl((P_{1,1}+P_{2,1})(P_{3,3}+P_{3,4})\bigr)&=1.
\label{eq:projector-criterion1}
\end{align}
The completion is then unique, and for $r,c\in\{1,2\}$
\begin{equation}\label{eq:projector-reconstruction}
\begin{aligned}
P_{r,2+c}&=(P_{r,1}+P_{r,2})^\perp\,(P_{3,2+c}+P_{4,2+c})^\perp,\\
P_{2+r,c}&=(P_{2+r,3}+P_{2+r,4})^\perp\,(P_{1,c}+P_{2,c})^\perp.
\end{aligned}
\end{equation}
\end{theorem}
In particular, the first and the fourth block determine the grid.

The reconstruction theorem admits a quantum-logical interpretation~\cite{BvN}: each missing block is the unique common refinement of two compatible binary projective decompositions, with all four joint outcomes of rank one.

Let $\mathcal L=\mathcal P(\mathbb C^4)$ be the orthomodular
lattice of orthogonal projections, with meet $\wedge$,
join $\vee$, and orthocomplementation ${}^\perp$. In quantum logic, orthogonal projections represent sharp
propositions, and an element $P\in\mathcal L$ is an atom if $P\ne0$ and
every $Q\in\mathcal L$ with $Q\le P$ satisfies $Q=0$ or $Q=P$.
In $\mathcal P(\mathbb C^4)$, the atoms are precisely the
rank-one projections.
Two elements $X,Y\in\mathcal L$ are called compatible if
$X=(X\wedge Y)\vee(X\wedge Y^\perp)$.
For orthogonal projections, compatibility is equivalent to
$XY=YX$; in this case $X\wedge Y=XY$.
Two elements are compatible if and only if they belong to a
common Boolean subalgebra of $\mathcal L$. In this sense,
compatible propositions admit a common classical logical
description. A SudoQ $\mathbf P$ is an array of atoms such that
every row, column and block is an orthogonal decomposition
of $I_4$, the greatest element of $\mathcal L$.

This is the quantum-logic version of the reconstruction theorem.
\begin{theorem}[Lattice-theoretic reconstruction]
\label{thm:lattice-reconstruction}
The partial grid~\eqref{eq:partial-grid} admits a SudoQ completion if
and only if
\begin{enumerate}[label=(\roman*)]
\item $P_{1,1}\vee P_{1,2}$ and $P_{3,3}\vee P_{4,3}$ are compatible, and
      $(P_{1,1}\vee P_{1,2})\wedge (P_{3,3}\vee P_{4,3})$ is an atom; %of $\mathcal L$;
\item $P_{1,1}\vee P_{2,1}$ and $P_{3,3}\vee P_{3,4}$ are compatible, and
      $(P_{1,1}\vee P_{2,1})\wedge (P_{3,3}\vee P_{3,4})$ is an atom. %of $\mathcal L$.
\end{enumerate}
The completion is then unique, and for $r,c\in\{1,2\}$
\begin{equation}
    \label{eq:projector-reconstruction2}
    \begin{aligned}
P_{r,2+c}&=(P_{r,1}\vee P_{r,2})^\perp\wedge (P_{3,2+c}\vee P_{4,2+c})^\perp,\\
P_{2+r,c}&=(P_{2+r,3}\vee P_{2+r,4})^\perp\wedge (P_{1,c}\vee P_{2,c})^\perp.
\end{aligned}
\end{equation}
\end{theorem}

\section{Cardinality and the overlap invariant}\label{sec:cardinality}
\begin{theorem}[Cardinality through opposite-block overlaps]
\label{th:cardinality}
For every SudoQ $\mathbf P$ of size two,
\begin{equation}\label{eq:intrinsic-cardinality}
c(\mathbf P)
=\#\{(a,b,r,c)\in\{1,2\}^4:\tr(P_{a,b}P_{2+r,2+c})>0\},
\end{equation}
and $c(\mathbf P) \in \{4,6,8,16\}$.
\end{theorem}

Examples~\ref{ex:projector-cardinalities} and~\ref{ex:product} show
that all four cases occur.

\begin{example}[Classical grids]\label{ex:classical}
The classical Sudoku on the left gives the classical SudoQ on the right:
\[
\left[\begin{array}{cc|cc}
1&2&4&3\\3&4&2&1\\\hline4&3&1&2\\2&1&3&4
\end{array}\right]
\qquad\longmapsto\qquad
\left[\begin{array}{cc|cc}
E_1&E_2&E_4&E_3\\E_3&E_4&E_2&E_1\\\hline
E_4&E_3&E_1&E_2\\E_2&E_1&E_3&E_4
\end{array}\right],
\]
of cardinality four.
\end{example}

\begin{example}[Cardinalities four, six and eight]
\label{ex:projector-cardinalities}
Let $P_{1},P_{2},P_{3},P_{4}$ be mutually orthogonal rank-one
projections, and split $P_{1}+P_{2}=R_1+R_2$ and
$P_{3}+P_{4}=R_3+R_4$ into rank-one projections. Then
\[
\mathbf P=\left[\begin{array}{cc|cc}
P_{1}&P_{2}&R_4&R_3\\
P_{3}&P_{4}&R_2&R_1\\\hline
P_{4}&P_{3}&R_1&R_2\\
P_{2}&P_{1}&R_3&R_4
\end{array}\right]
\]
is a SudoQ: every row, column and block sums to $\sum_{j=1}^{4}P_{j}$
or to $\sum_{j=1}^{4}R_j$, both equal to $I_4$. If
$R_1\in\{P_{1},P_{2}\}$, then $R_2=P_{1}+P_{2}-R_1$ is the
other one; and $R_1,R_2$ are orthogonal to
$P_{3},P_{4}$. Hence each of the pairs
$\{R_1,R_2\}$, $\{R_3,R_4\}$ either coincides with the corresponding
pair of the first block or adds two new projections, and
$c(\mathbf P)=4,6,8$ according as none, one or both pairs are new.
\end{example}

\begin{example}[Product grids]\label{ex:product}
Write $\C^4=\C^2\otimes\C^2$, so that $e_{2(a-1)+b}=e_a\otimes e_b$, $a,b \in \{1,2\}$
and choose rank-one projections on $\C^2$ with
$Q_1+Q_2=\widetilde Q_1+\widetilde Q_2=L_1+L_2
=\widetilde L_1+\widetilde L_2=I_2$. Then
\[
\mathbf P=\left[\begin{array}{cc|cc}
Q_1\otimes L_1&Q_1\otimes L_2&Q_2\otimes\widetilde L_2&Q_2\otimes\widetilde L_1\\
Q_2\otimes L_1&Q_2\otimes L_2&Q_1\otimes\widetilde L_2&Q_1\otimes\widetilde L_1\\\hline
\rule{0pt}{2.5ex}
\widetilde Q_2\otimes L_2&\widetilde Q_2\otimes L_1&\widetilde Q_1\otimes\widetilde L_1&\widetilde Q_1\otimes\widetilde L_2\\
\widetilde Q_1\otimes L_2&\widetilde Q_1\otimes L_1&\widetilde Q_2\otimes\widetilde L_1&\widetilde Q_2\otimes\widetilde L_2
\end{array}\right]
\]
is a SudoQ, as every row, column and block sums to $I_4=I_2\otimes I_2$.
On $\C^2$ the pairs $\{Q_1,Q_2\}$ and $\{\widetilde Q_1,\widetilde Q_2\}$
are either equal or disjoint, and likewise for the $L$'s. The entries
of $\mathbf P$ are all the products $Q\otimes L$ of a $Q$ and an $L$
occurring in the grid, so $c(\mathbf P)=4,8,16$ according as both,
one or neither of the equalities
$\{\widetilde Q_1,\widetilde Q_2\}=\{Q_1,Q_2\}$,
$\{\widetilde L_1,\widetilde L_2\}=\{L_1,L_2\}$ hold. 
\end{example}

The labelled overlaps carry much more
information than their number.
In terms of quantum logic, the number of distinct atomic propositions in the grid equals the number of nonorthogonal pairs formed by taking one atom from each of the two opposite blocks.

\begin{theorem}[Complete overlap invariant]\label{cor:complete-overlaps}
Two SudoQ $\mathbf P$, $\mathbf P'$ of size two are unitarily
equivalent if and only if
\[
\tr(P_{a,b}P_{2+r,2+c})=\tr(P'_{a,b}P'_{2+r,2+c})
\qquad\text{for all }a,b,r,c\in\{1,2\}.
\]
\end{theorem}

The overlaps do not record relative phases; the Sudoku
constraints force exactly the phase relations needed for a single
unitary to match the two grids. In the language of
Section~\ref{sec:fractional}, $\tr(P_{a,b}P_{2+r,2+c})$ is the
probability of the outcome $(a,b)$ when the fourth block is measured
in the basis of the first, so Theorem~\ref{cor:complete-overlaps} says
that this measurement determines the grid up to unitary equivalence; we return to this in
Section~\ref{sec:realparam}. As an immediate consequence of Theorem~\ref{cor:complete-overlaps}, we obtain the following corollary.

\begin{corollary}\label{cor:real}
Every SudoQ $\mathbf{P}$ of size two is unitarily equivalent to a real one, i.e. there exists an orthonormal basis $\{f_a\}_{a=1}^4$ of $\C^4$ such that $\langle f_a,P_{r,c} f_b\rangle$ is real for all $a,b,r,c\in\{1,2,3,4\}$. 
\end{corollary}

\section{Matrix representation}\label{sec:matrix-representation}
Choosing a unit vector in the range of each projection, and arranging
the four vectors of a block as the columns of a unitary matrix, turns
the description by projections into a parametrisation by matrices.

\begin{definition}\label{defn:fourmat}
Let $\mathbf P$ be a SudoQ of size two. Matrices $R,U,V,W\in\U(4)$
\emph{represent} its four blocks if
\begin{equation}\label{eq:projector-blocks}
\begin{aligned}
P_{r,c}&=R E_{2(r-1)+c}R^*,&
P_{r,2+c}&=U E_{2(r-1)+c}U^*,\\
P_{2+r,c}&=V E_{2(r-1)+c}V^*,&
P_{2+r,2+c}&=W E_{2(r-1)+c}W^*,
\end{aligned}\qquad r,c\in\{1,2\},
\end{equation}
and we write
\begin{equation}\label{eq:block-representation}
\mathbf P\longleftrightarrow
\begin{array}{c|c}R&U\\\hline V&W\end{array}.
\end{equation}
\end{definition}

Each representative is unique up to right multiplication by a
diagonal unitary. Any four unitaries define an array satisfying the
block conditions, but in general not the row and column conditions.

\begin{proposition}[Normalisation and residual equivalence]
\label{prop:normalization}
Every SudoQ $\mathbf P\longleftrightarrow
\begin{array}{c|c}R&U\\\hline V&W\end{array}$ is unitarily equivalent
to the \emph{normalised} SudoQ
$\begin{array}{c|c}I_4&R^*U\\\hline R^*V&R^*W\end{array}$.
Two normalised SudoQ
$\begin{array}{c|c}I_4&U\\\hline V&W\end{array}$ and
$\begin{array}{c|c}I_4&U'\\\hline V'&W'\end{array}$
are unitarily equivalent if and only if there are diagonal unitaries
% $\Delta,\Delta_U,\Delta_V,\Delta_W$ with
% \begin{equation}\label{eq:residual-equivalence}
% U'=\Delta U\Delta_U,\qquad V'=\Delta V\Delta_V,\qquad W'=\Delta W\Delta_W.
% \end{equation}
$D_L,D_U,D_V,D_W$ with
\begin{equation}\label{eq:residual-equivalence}
U'=D_L U D_U,\qquad V'=D_L V D_V,\qquad W'=D_L W D_W.
\end{equation}
\end{proposition}

The right factors rephase the columns; the common left factor $D_L$
rephases the canonical basis, which leaves every $E_j$ fixed. The
fourth block of a normalised grid is therefore a point of the
following double coset space.

\begin{definition}\label{def:matrix-quotients}
Let $\mathbb T_n\subseteq\U(n)$ be the diagonal unitaries. For
$M,M'\in\U(n)$ write $M\sim M'$ if $M'=D_LMD_R$ with $D_L,D_R\in\mathbb T_n$,
and let $\mathcal B_n=\U(n)/{\sim}$, with classes $[M]_\sim$.
\end{definition}

We shall also use
\begin{equation}\label{eq:swap}
\mathscr S(M)=KMK,\qquad
K=\begin{pmatrix}1&0&0&0\\0&0&1&0\\0&1&0&0\\0&0&0&1\end{pmatrix},
\end{equation}
which exchanges the coordinates $2$ and $3$; since $K\mathbb T_4K=\mathbb T_4$, it is
an involution of $\mathcal B_4$. Admissibility of $W$ is expressed
through the $2\times2$ submatrices $W_{[r,c]}$ (rows $\{2r-1,2r\}$, columns $\{c,c+2\}$) and $W^{[r,c]}$ (rows $\{r,r+2\}$, columns $\{2c-1,2c\}$), $r,c\in\{1,2\}$,
\begin{equation}\label{eq:compressed-matrices}
W_{[r,c]}=
\begin{pmatrix}w_{2r-1,c}&w_{2r-1,c+2}\\
w_{2r,c}&w_{2r,c+2}\end{pmatrix},\qquad
 W^{[r,c]}=
\begin{pmatrix}w_{r,2c-1}&w_{r,2c}\\
w_{r+2,2c-1}&w_{r+2,2c}\end{pmatrix}.
\end{equation}

\begin{theorem}[Matrix criterion and parametrisation]
\label{cor:matrix-reconstruction}\label{prop:quotient-reconstruction}
Let $[W]_{\sim}\in\mathcal B_4$. The partial grid
\begin{equation}\label{eq:canS}
\begin{array}{c|c}I_4& ? \\\hline ? &W\end{array}
\end{equation}
admits a SudoQ completion if and only if
$\det W_{[r,c]}=\det W^{[r,c]}=0$ for all $r,c\in\{1,2\}$, that is,
\begin{equation}\label{systemW}
\begin{aligned}
w_{11}w_{23}&=w_{13}w_{21},& w_{11}w_{32}&=w_{31}w_{12},&
w_{22}w_{14}&=w_{12}w_{24},& w_{22}w_{41}&=w_{21}w_{42},\\
w_{33}w_{14}&=w_{13}w_{34},& w_{33}w_{41}&=w_{31}w_{43},&
w_{44}w_{23}&=w_{24}w_{43},& w_{44}w_{32}&=w_{42}w_{34}.
\end{aligned}
\end{equation}
The completion is then unique, and the admissible classes in
$\mathcal B_4$ are in bijection with the unitary equivalence classes
of SudoQ of size two.
\end{theorem}

Here uniqueness refers to the projections in the two missing
blocks. For a fixed representative $W$, the matrices $U$ and $V$
representing these blocks are determined up to independent right
multiplication by diagonal unitaries, which only changes the phases
of their columns.

\subsection{Overlaps, cardinality and classical Sudoku}\label{sec:matrix-cardinality}
Consider a normalised SudoQ with block representation
\[
\mathbf P\longleftrightarrow
\begin{array}{c|c}
I_4&U\\\hline
V&W
\end{array},
\qquad U,V,W\in\U(4).
\]
Each matrix represents the four projections in the corresponding
block through its columns, as in~\eqref{eq:projector-blocks}.
The overlaps between the first and fourth blocks are
\[
\tr(P_{a,b}P_{2+r,2+c})
=\tr(E_{2(a-1)+b}WE_{2(r-1)+c}W^*)
=|w_{2(a-1)+b,\,2(r-1)+c}|^2.
\]
The representative $W$ is defined up to the equivalence
$W\sim W'=D_LWD_R$ of Definition~\ref{def:matrix-quotients},
where $D_L,D_R$ are diagonal unitaries. Such rephasing preserves
$|w_{ij}|^2$ and hence the positions and number of nonzero entries.
Thus both the overlaps and the count below depend only on the
class $[W]_\sim$, not on the chosen representative.
By Theorem~\ref{th:cardinality}, the cardinality is therefore the
number of nonzero entries of $W=(w_{ij})_{i,j=1}^{4}$:
\[
c(\mathbf P)=\#\{(i,j)\in\{1,2,3,4\}^2:w_{ij}\neq0\}
\in\{4,6,8,16\}.
\]

The classical grids are precisely those for which $W$ is equivalent,
under diagonal rephasing, to an admissible permutation matrix.
Indeed, a unitary matrix has exactly four nonzero entries if and only
if it has one entry of modulus one in each row and column; rephasing
turns these entries into ones. Each such class contains a unique
permutation matrix. Exactly twelve of the $24$ permutation matrices
satisfy~\eqref{systemW}; their completed grids are listed in
Table~\ref{tab:classes}. 
With the first block fixed as
$\left[\begin{smallmatrix}1&2\\3&4\end{smallmatrix}\right]$,
there are twelve classical completions. In an arbitrary classical
Sudoku, the first block contains the four symbols exactly once, so
it can be arranged in $4!$ ways. For each arrangement there is a
unique permutation of the symbols that sends it to the fixed first
block. Applying this permutation to the whole grid gives a bijection
between its completions and the twelve completions of the fixed
block. Hence there are $4!\cdot12=288$ classical Sudoku of size two.

\FloatBarrier

\subsection{The parameter space}
\label{sec:parameter_space}
By Theorem~\ref{prop:quotient-reconstruction}, the space of unitary
equivalence classes of SudoQ of size two is
\begin{equation}\label{eq:space-W}
\mathcal W=\left\{[W]_{\sim}\in\mathcal B_4:
W\text{ satisfies }\eqref{systemW}\right\},
\end{equation}
with the topology induced by $\mathcal B_4$. From now on we drop the
brackets: a matrix stands for its class, and equalities in
$\mathcal W$ are equalities of classes.

Every class in $\mathcal B_2$ has a unique representative
\begin{equation}\label{eq:A}
A(\alpha)=\begin{pmatrix}
\cos(\alpha)&\sin(\alpha)\\-\sin(\alpha)&\cos(\alpha)
\end{pmatrix},\qquad\alpha\in[0,\pi/2],
\end{equation}
since rephasing rows and columns makes a $2\times2$ unitary real of
this form, $\cos(\alpha)$ being the modulus of its upper-left entry. Note that $A(0)=I_2$ and
$A(\pi/2)=J=\left(\begin{smallmatrix}0&1\\-1&0\end{smallmatrix}\right)$.
We now introduce five families of matrices: for $\alpha,\beta\in[0,\pi/2]$ let
\begin{equation}\label{eq:angle-representatives}
\begin{aligned}
T(\alpha,\beta)&=A(\alpha)\otimes A(\beta),\\
D(\alpha,\beta)&=\begin{pmatrix}A(\alpha)&0\\0&A(\beta)\end{pmatrix},&
O(\alpha,\beta)&=\begin{pmatrix}0&A(\alpha)\\A(\beta)&0\end{pmatrix},\\
C_{\mathrm{w}}(\alpha,\beta)&=K D(\alpha,\beta)K,&
C_{\mathrm{b}}(\alpha,\beta)&=K O(\alpha,\beta)K,
\end{aligned}
\end{equation}
and $\mathcal T = \{T(\alpha,\beta): \alpha,\beta\in[0,\pi/2] \}$,
$$
\mathcal D = \{D(\alpha,\beta): \alpha,\beta\in[0,\pi/2] \}, \quad \mathcal O = \{O(\alpha,\beta): \alpha,\beta\in[0,\pi/2] \}, 
$$
$$
\mathcal C_{\mathrm w} = \{C_{\mathrm{w}}(\alpha,\beta): \alpha,\beta\in[0,\pi/2] \}, \quad \mathcal C_{\mathrm b} = \{C_{\mathrm{b}}(\alpha,\beta): \alpha,\beta\in[0,\pi/2] \}. 
$$
The four
\emph{sparse families} have the supports
\begin{equation}\label{eq:support-patterns}
\begin{array}{c@{\qquad}c@{\qquad}c@{\qquad}c}
\mathcal D&\mathcal O&\mathcal C_{\mathrm{w}}&\mathcal C_{\mathrm{b}}\\[2pt]
\begin{pmatrix}*&*&0&0\\ *&*&0&0\\0&0&*&*\\0&0&*&*\end{pmatrix}&
\begin{pmatrix}0&0&*&*\\0&0&*&*\\ *&*&0&0\\ *&*&0&0\end{pmatrix}&
\begin{pmatrix}*&0&*&0\\0&*&0&*\\ *&0&*&0\\0&*&0&*\end{pmatrix}&
\begin{pmatrix}0&*&0&*\\ *&0&*&0\\0&*&0&*\\ *&0&*&0\end{pmatrix}
\end{array},
\end{equation}
where $*$ marks an entry that may be nonzero; the subscripts refer to
the white ($i+j$ even) and black ($i+j$ odd) squares of a
checkerboard. A matrix, or a class, is \emph{full} if it has no zero
entries; $\mathcal T$ is the \emph{tensor family} and all the entries of $T(\alpha, \beta)$ are nonzero for
$\alpha,\beta\in(0,\pi/2)$. The following theorem shows that these families are contained in \(\mathcal W\) and together exhaust it.

\begin{theorem}[The five-square model]\label{setW}
Every $W\in\mathcal W$ can be written as
\begin{equation}\label{normal_form}
W=X(\alpha,\beta),\qquad X\in\{T,D,O,C_{\mathrm{w}},C_{\mathrm{b}}\},
\quad\alpha,\beta\in[0,\pi/2],
\end{equation}
and for a given $X$ the angles $\alpha,\beta$ are unique. Thus
$\mathcal W=\mathcal T\cup\mathcal D\cup\mathcal O
\cup\mathcal C_{\mathrm{w}}\cup\mathcal C_{\mathrm{b}}$, and
$\mathcal W$ is homeomorphic to the quotient of five copies of
$[0,\pi/2]^2$ by the identifications generated by
\begin{equation}\label{eq:four-gluings}
\begin{aligned}
T(0,\beta)&=D(\beta,\beta),&
T(\pi/2,\beta)&=O(\beta,\beta),\\
T(\alpha,0)&=C_{\mathrm{w}}(\alpha,\alpha),&
T(\alpha,\pi/2)&=C_{\mathrm{b}}(\alpha,\alpha),
\end{aligned}\qquad \alpha,\beta\in[0,\pi/2].
\end{equation}
These are exactly the intersections
of $\mathcal T$ with the sparse families, and the pairwise
intersections of the sparse families are
\begin{equation}\label{eq:sparse-intersections}
\begin{aligned}
\mathcal D\cap\mathcal O&=\varnothing,
&\mathcal C_{\mathrm{w}}\cap\mathcal C_{\mathrm{b}}&=\varnothing,\\
\mathcal D\cap\mathcal C_{\mathrm{w}}&=\{T(0,0)\},
&\mathcal D\cap\mathcal C_{\mathrm{b}}&=\{T(0,\pi/2)\},\\
\mathcal O\cap\mathcal C_{\mathrm{w}}&=\{T(\pi/2,0)\},
&\mathcal O\cap\mathcal C_{\mathrm{b}}&=\{T(\pi/2,\pi/2)\}.
\end{aligned}
\end{equation}
In particular, the nonempty intersections of three families (including $\mathcal T$) are the
four corners of $\mathcal T$, and any four families have empty
intersection. Finally, $\mathscr S(\mathcal W)=\mathcal W$.
\end{theorem}
The classical classes are precisely the corners of the five squares.
For $X\in\{T,D,O,C_{\mathrm{w}},C_{\mathrm{b}}\}$ and $i,j\in\{0,1\}$, denote by $S^X_{ij}$ the classical Sudoku
corresponding to $X(i\pi/2,j\pi/2)$, with the first block fixed in
canonical order. The identifications~\eqref{eq:four-gluings} leave
twelve distinct corners, listed in Table~\ref{tab:classes}: the four
corners of $\mathcal T$ are of Type~1, and the eight off-diagonal
corners of the sparse squares are of Type~2 (see Section~\ref{sec:disc}).

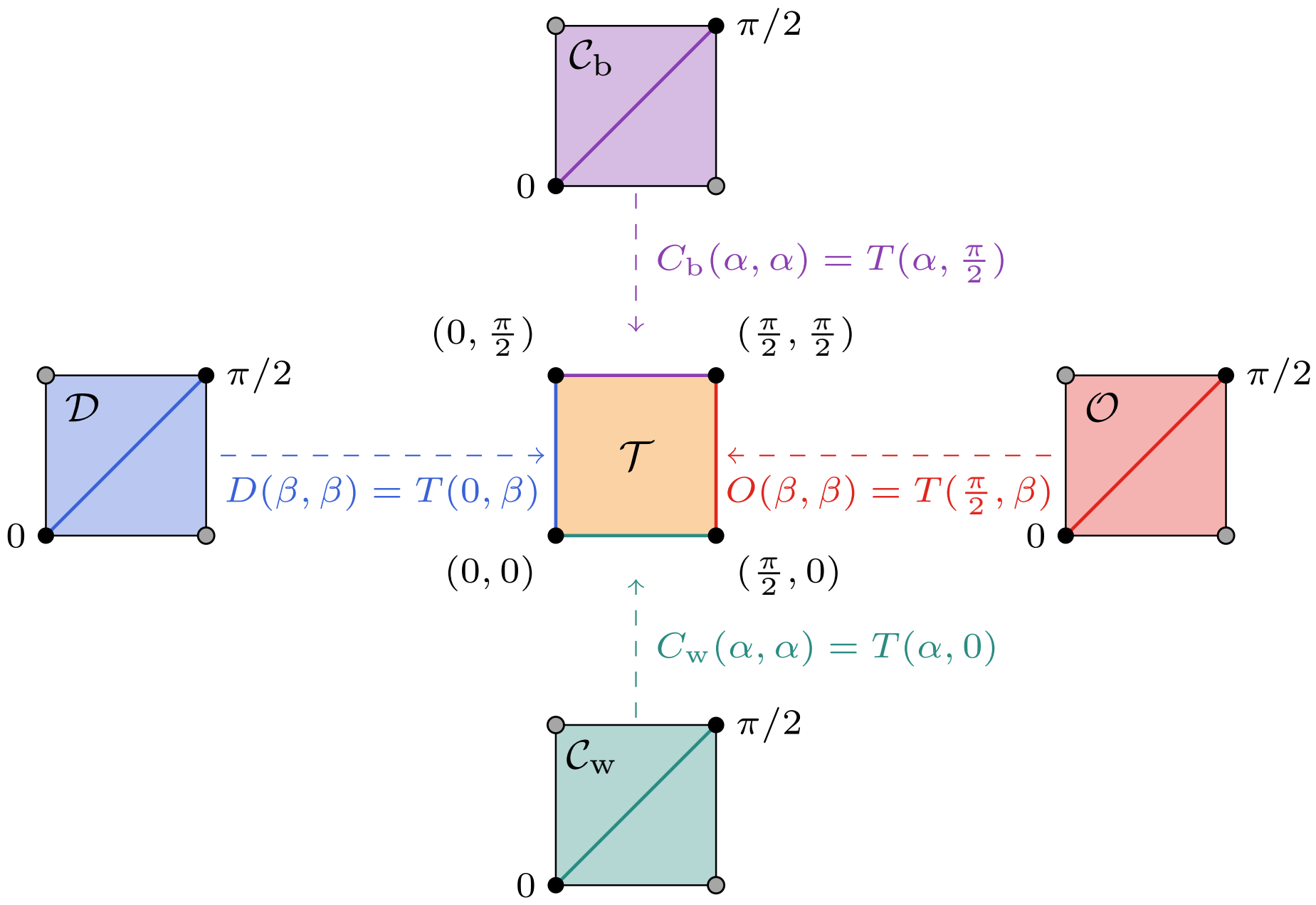
\begin{figure}[thbp]
\centering
\resizebox{0.80\linewidth}{!}{%
\begin{tikzpicture}[x=1cm,y=1cm,font=\scriptsize]
\def\a{1.1}
% tensor square
\fill[cT!45] (0,0) rectangle (\a,\a);
\draw[cD,line width=0.65pt] (0,0)--(0,\a); \draw[cO,line width=0.65pt] (\a,0)--(\a,\a);
\draw[cCw,line width=0.65pt] (0,0)--(\a,0); \draw[cCb,line width=0.65pt] (0,\a)--(\a,\a);
\foreach \p in {(0,0),(\a,0),(0,\a),(\a,\a)} \fill \p circle (1.6pt);
\node at (\a/2,\a/2) {$\mathcal T$};
\node[below left] at (0,0) {\tiny$(0,0)$}; \node[below right] at (\a,0) {\tiny$(\frac\pi2,0)$};
\node[above left] at (0,\a) {\tiny$(0,\frac\pi2)$}; \node[above right] at (\a,\a) {\tiny$(\frac\pi2,\frac\pi2)$};
% sparse squares
\fsq{-3.5}{0}{cD}{$\mathcal D$}
\fsq{3.5}{0}{cO}{$\mathcal O$}
\fsq{0}{2.4}{cCb}{$\mathcal C_{\mathrm b}$}
\fsq{0}{-2.4}{cCw}{$\mathcal C_{\mathrm w}$}
% dashed arrows
\draw[cD,dashed,->,line width=0.3pt] (-2.3,\a/2) -- (-0.08,\a/2) node[midway,below,font=\tiny] {$D(\beta,\beta)=T(0,\beta)$};
\draw[cO,dashed,->,line width=0.3pt] (3.4,\a/2) -- (\a+0.08,\a/2) node[midway,below,font=\tiny] {$O(\beta,\beta)=T(\frac\pi2,\beta)$};
\draw[cCb,dashed,->,line width=0.3pt] (\a/2,2.35) -- (\a/2,\a+0.3) node[midway,right,font=\tiny] {$C_{\rm b}(\alpha,\alpha)=T(\alpha,\frac\pi2)$};
\draw[cCw,dashed,->,line width=0.3pt] (\a/2,-1.25) -- (\a/2,-0.3) node[midway,right,font=\tiny] {$C_{\rm w}(\alpha,\alpha)=T(\alpha,0)$};
\pgfresetboundingbox\path[use as bounding box] (-4.0,-2.9) rectangle (5.1,3.9);
\end{tikzpicture}}

\caption{The five-square model. In each square
the parameter domain is $[0,\pi/2]^2$.
Each coloured diagonal of a sparse square is identified pointwise with
the side of the same colour of the tensor square $\mathcal T$, in the
parameter order of~\eqref{eq:four-gluings}. Circles mark the twelve
classical Sudoku, black for Type~1 and grey for Type~2
(Section~\ref{sec:disc}). The four corners of $\mathcal T$ are the
only points lying in two sparse families at once: they are the
Type~1 grids, each drawn three times, while the off-diagonal corners
of the sparse squares are the eight Type~2 grids.}
\label{fig:five-squares}
\end{figure}

\begin{table}[htbp]
\centering
\renewcommand{\arraystretch}{1.15}
\begin{tabular}{cccc}
\multicolumn{4}{c}{\small Type 1}\\[3pt]
\shid{1&2&4&3}{3&4&2&1}{4&3&1&2}{2&1&3&4} &
\shid{1&2&3&4}{3&4&1&2}{4&3&2&1}{2&1&4&3} &
\shid{1&2&4&3}{3&4&2&1}{2&1&3&4}{4&3&1&2} &
\shid{1&2&3&4}{3&4&1&2}{2&1&4&3}{4&3&2&1}\\
{\scriptsize $S^T_{00}=S^D_{00}=S^{C_{\mathrm w}}_{00}$} &
{\scriptsize $S^T_{01}=S^D_{11}=S^{C_{\mathrm b}}_{00}$} &
{\scriptsize $S^T_{10}=S^O_{00}=S^{C_{\mathrm w}}_{11}$} &
{\scriptsize $S^T_{11}=S^O_{11}=S^{C_{\mathrm b}}_{11}$}\\[10pt]
\multicolumn{4}{c}{\small Type 2}\\[3pt]
\shid{1&2&3&4}{3&4&2&1}{4&3&1&2}{2&1&4&3} &
\shid{1&2&4&3}{3&4&1&2}{4&3&2&1}{2&1&3&4} &
\shid{1&2&3&4}{3&4&2&1}{2&1&4&3}{4&3&1&2} &
\shid{1&2&4&3}{3&4&1&2}{2&1&3&4}{4&3&2&1}\\
{\scriptsize $S^D_{01}$} & {\scriptsize $S^D_{10}$} &
{\scriptsize $S^O_{01}$} & {\scriptsize $S^O_{10}$}\\[6pt]
\shid{1&2&4&3}{3&4&2&1}{2&3&1&4}{4&1&3&2} &
\shid{1&2&4&3}{3&4&2&1}{4&1&3&2}{2&3&1&4} &
\shid{1&2&3&4}{3&4&1&2}{2&3&4&1}{4&1&2&3} &
\shid{1&2&3&4}{3&4&1&2}{4&1&2&3}{2&3&4&1}\\
{\scriptsize $S^{C_{\mathrm w}}_{01}$} & {\scriptsize $S^{C_{\mathrm w}}_{10}$} &
{\scriptsize $S^{C_{\mathrm b}}_{01}$} & {\scriptsize $S^{C_{\mathrm b}}_{10}$}
\end{tabular}
\caption[The twelve classical Sudoku of size two]{The twelve classical Sudoku of size two with first block
$\left[\begin{smallmatrix}1&2\\3&4\end{smallmatrix}\right]$, the
symbol $i$ standing for $E_i$. The grid $S^X_{ij}$ corresponds to
$X(i\pi/2,j\pi/2)$. Each Type~1 grid is a corner of the tensor square
and of two sparse squares, whereas each Type~2 grid is a corner of
one sparse square.}
\label{tab:classes}
\end{table}
For an illustration of Theorem~\ref{setW}, see Figure~\ref{fig:five-squares}. 

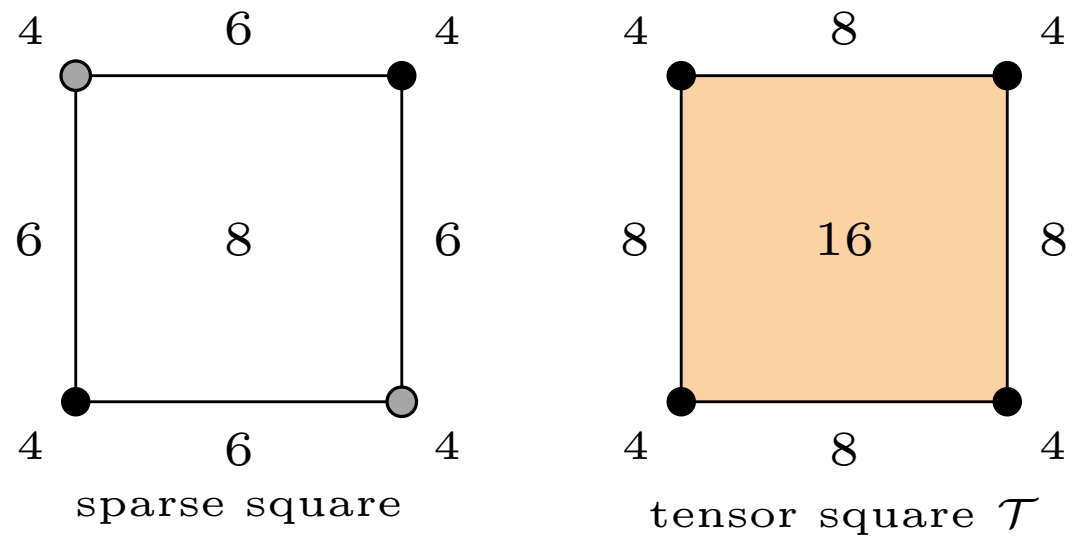
\begin{figure}[htbp]
\centering
\resizebox{0.60\linewidth}{!}{%
\begin{tikzpicture}[x=1cm,y=1cm,font=\fontsize{6}{7}\selectfont]
% sparse square
\fill[white] (0,0) rectangle (1.4,1.4); \draw[line width=0.35pt] (0,0) rectangle (1.4,1.4);
\node at (0.7,0.7) {$8$};
\node[left] at (0,0.7) {$6$}; \node[right] at (1.4,0.7) {$6$}; \node[below] at (0.7,0) {$6$}; \node[above] at (0.7,1.4) {$6$};
\foreach \p/\q in {0/0,1.4/1.4} \fill (\p,\q) circle (1.8pt);
\foreach \p/\q in {1.4/0,0/1.4} \filldraw[fill=black!35] (\p,\q) circle (1.8pt);
\node[below left] at (0,0) {\fontsize{5}{6}\selectfont $4$}; \node[above right] at (1.4,1.4) {\fontsize{5}{6}\selectfont $4$};
\node[below right] at (1.4,0) {\fontsize{5}{6}\selectfont $4$}; \node[above left] at (0,1.4) {\fontsize{5}{6}\selectfont $4$};
\node[below=8pt,align=center,font=\fontsize{5}{6}\selectfont] at (0.7,0) {sparse square};
% tensor square
\begin{scope}[xshift=2.6cm]
\fill[cT!45] (0,0) rectangle (1.4,1.4); \draw[line width=0.35pt] (0,0) rectangle (1.4,1.4);
\node at (0.7,0.7) {$16$};
\node[left] at (0,0.7) {$8$}; \node[right] at (1.4,0.7) {$8$}; \node[below] at (0.7,0) {$8$}; \node[above] at (0.7,1.4) {$8$};
\foreach \p/\q in {0/0,1.4/0,0/1.4,1.4/1.4} \fill (\p,\q) circle (1.8pt);
\node[below left] at (0,0) {\fontsize{5}{6}\selectfont $4$}; \node[above right] at (1.4,1.4) {\fontsize{5}{6}\selectfont $4$};
\node[below right] at (1.4,0) {\fontsize{5}{6}\selectfont $4$}; \node[above left] at (0,1.4) {\fontsize{5}{6}\selectfont $4$};
\node[below=8pt] at (0.7,0) {\fontsize{5}{6}\selectfont  tensor square $\mathcal T$};
\end{scope}
\pgfresetboundingbox\path[use as bounding box] (-1.0,-0.85) rectangle (5.0,1.75);
\end{tikzpicture}}

\caption{Cardinality of the SudoQ classes in the parameter squares of
Theorem~\ref{setW}. In a sparse square the cardinality is eight in the
interior, six on the four open edges and four at the corners; in the
tensor square it is sixteen in the interior, while the boundary is
glued to the sparse diagonals by~\eqref{eq:four-gluings} and carries
cardinality eight on the open sides and four at the corners. The
twelve corners which remain distinct after the identifications are
exactly the twelve classical Sudoku of Table~\ref{tab:classes}; as in
Figure~\ref{fig:five-squares}, black circles mark those of Type~1,
which in a sparse square are the two corners on the glued diagonal,
and grey circles those of Type~2.}
\label{fig:cardinality}
\end{figure}

Counting the nonzero entries of the representatives in
\eqref{eq:angle-representatives} gives the cardinalities shown in
Figure~\ref{fig:cardinality}: in each sparse square they are eight
in the interior, six on the open edges and four at the corners;
in the tensor square they are sixteen in the interior, eight on
the open sides and four at the corners.

\FloatBarrier

\subsection{Reconstruction in angular coordinates}\label{sec:angular-reconstruction}
For $R=I_4$ the first block is canonical, and for $W=X(\alpha,\beta)$, $X \in \{T,D,O, C_{\mathrm{w}}, C_{\mathrm{b}}\}$, $\alpha, \beta \in [0,\pi/2]$,
Theorem~\ref{thm:WUV} reads
\begin{equation}\label{eq:angular-reconstruction-projections}
\begin{aligned}
UE_{2(r-1)+c}U^*&=\bigl(I_4-E_{2r-1}-E_{2r}\bigr)
\bigl(I_4-W(E_c+E_{c+2})W^*\bigr),\\
VE_{2(r-1)+c}V^*&=\bigl(I_4-W(E_{2r-1}+E_{2r})W^*\bigr)
\bigl(I_4-E_c-E_{c+2}\bigr).
\end{aligned}
\end{equation}
Choosing unit vectors in these ranges gives
\begin{equation}\label{eq:angular-blocks}
\renewcommand{\arraystretch}{1.4}
\begin{array}{c|c|c}
W&U&V\\\hline
T(\alpha,\beta)
 &J\otimes\bigl(A(\beta)J\bigr)
 &\bigl(A(\alpha)J\bigr)\otimes J\\
D(\alpha,\beta)
 &O(\alpha,\beta)(I_2\otimes J)&J\otimes J\\
O(\alpha,\beta)
 &O(\alpha,\beta)(I_2\otimes J)&I_2\otimes J\\
C_{\mathrm{w}}(\alpha,\beta)
 &J\otimes J&\mathscr S\!\left(O(\alpha,\beta)(I_2\otimes J)\right)\\
C_{\mathrm{b}}(\alpha,\beta)
 &J\otimes I_2&\mathscr S\!\left(O(\alpha,\beta)(I_2\otimes J)\right)
\end{array}
\end{equation}

\begin{remark}[Comparison with the vector parametrisation]\label{rmk:vector}
The vector parametrisation of $4\times4$ SudoQ in~\cite{KZ} and the
normal forms above describe the same grids, but the latter organise
them into unitary equivalence classes. The passage between the two
descriptions is explicit: if $u_{r,c}$ is a unit vector in a cell,
then $P_{r,c}=u_{r,c}u_{r,c}^*$; conversely, the columns of
$I_4,U,V,W$ in~\eqref{eq:angular-blocks} give vector representatives
of all sixteen cells of the normalised grid.

There are two distinct freedoms to account for. Multiplying each
$u_{r,c}$ by an independent phase leaves the projection in that cell
unchanged. Applying a single unitary to all sixteen vectors generally
changes the projections, but preserves the unitary equivalence class
of the grid. After the first block has been normalised, these freedoms
leave precisely the residual equivalence of
Proposition~\ref{prop:normalization}; for the fourth block it is
$W\sim D_LWD_R$, with diagonal unitaries $D_L,D_R$.

Thus the phases appearing in vector representatives should not be
counted automatically as parameters of inequivalent SudoQ.
Corollary~\ref{cor:real} shows that every class has a real
representative, and Theorem~\ref{setW} describes the remaining freedom
by a family $X$ and two angles $(\alpha,\beta)$. The angles are unique
within each family; the coincidences between different families are
exactly the identifications~\eqref{eq:four-gluings}. The five-square
model therefore records both the parameters of the equivalence
classes and how the families meet, including their classical limits.
\end{remark}

\section{Quotient by the classical cell symmetries}\label{sec:cell-symmetries}
We now consider the classical positional symmetries of Sudoku.
For a SudoQ of size two $\mathbf P=(P_{i,j})_{i,j=1}^{4}$, let $r$ exchange
the third and fourth rows, let $c$ exchange the third and fourth
columns, let $b$ exchange the two bands while preserving the order
of the rows within each band, and let $\tau$ transpose the grid.
Explicitly, setting $\sigma=(3\,4) \in \mathcal{S}_4$ and $\rho=(1\,3)(2\,4)\in \mathcal{S}_4$, we have for all $i,j\in\{1,2,3,4\}$:
\[
\begin{aligned}
(r\cdot\mathbf P)_{i,j}&=P_{\sigma(i),j},
&
(c\cdot\mathbf P)_{i,j}&=P_{i,\sigma(j)},\\
(b\cdot\mathbf P)_{i,j}&=P_{\rho(i),j},
&
(\tau\cdot\mathbf P)_{i,j}&=P_{j,i}.
\end{aligned}
\]
These operations permute the positions of the projections without
changing the projections themselves. They preserve the row, column
and block conditions and therefore map SudoQ to SudoQ.
They generate the classical positional symmetry group
\[
G_{\mathrm{cl}}=\langle r,c,\tau,b\rangle,
\qquad |G_{\mathrm{cl}}|=128.
\]
It is the full positional symmetry
group of the $4\times 4$ Sudoku~\cite{AFLT}; the relabelling of symbols is already
part of unitary equivalence. 
%Each $g\in G_{\mathrm{cl}}$ maps SudoQ to SudoQ. 
As for classical Sudoku, grids related by these symmetries and by a
relabelling are regarded as essentially the same.

\begin{definition}\label{defequiv_cl}
Two SudoQ $\mathbf P$, $\mathbf P'$ of size two are \emph{essentially
equivalent}, $\mathbf P\approx\mathbf P'$, if
$\mathbf P'\equiv g\cdot\mathbf P$ for some $g\in G_{\mathrm{cl}}$,
where $\equiv$ is the unitary equivalence of
Definition~\ref{defequiv}.
\end{definition}

\subsection{The induced action on the parameter space}
We now describe how the positional symmetries act on the matrix representation of a SudoQ and hence on its unitary equivalence class.
Let $\Pi=\left(\begin{smallmatrix}0&1\\1&0\end{smallmatrix}\right)$.
Right multiplication of a block representative by $\Pi\otimes I_2$,
$I_2\otimes\Pi$ and $K$ exchanges the two rows of the block, its two
columns, or transposes it. Renormalising the first block (by $K$ for
$\tau$, by $V^*$ for $b$), one finds that if
$\mathbf P\longleftrightarrow\begin{array}{c|c}I_4&U\\\hline V&W\end{array}$,
then
\begin{equation}\label{eq:normalized-generator-grids}
\begin{aligned}
r\cdot\mathbf P&\longleftrightarrow
\begin{array}{c|c}I_4&U\\\hline V(\Pi\otimes I_2)&W(\Pi\otimes I_2)\end{array},
&c\cdot\mathbf P&\longleftrightarrow
\begin{array}{c|c}I_4&U(I_2\otimes\Pi)\\\hline V&W(I_2\otimes\Pi)\end{array},\\[8pt]
\tau\cdot\mathbf P&\longleftrightarrow
\begin{array}{c|c}I_4&KVK\\\hline KUK&KWK\end{array},
&b\cdot\mathbf P&\longleftrightarrow
\begin{array}{c|c}I_4&V^*W\\\hline V^*&V^*U\end{array}.
\end{aligned}
\end{equation}

\begin{proposition}\label{lem:cell-action}
The generators of $G_{\mathrm{cl}}$ act on $\mathcal W$ by
\begin{equation}\label{eq:generator-fourth-blocks}
\begin{aligned}
r&:\ W\longmapsto W(\Pi\otimes I_2),&\qquad
c&:\ W\longmapsto W(I_2\otimes\Pi),\\
\tau&:\ W\longmapsto KWK,&\qquad
b&:\ W\longmapsto V^*U,
\end{aligned}
\end{equation}
where $U,V$ are given by~\eqref{eq:angular-blocks}. In coordinates,
\begin{equation}\label{eq:cell-action-table}
\setlength{\arraycolsep}{3pt}
\renewcommand{\arraystretch}{1.2}
\begin{array}{c|cccc}
W &r&c&\tau&b\\\hline
T(\alpha,\beta)&T(\pi/2-\alpha,\beta)&T(\alpha,\pi/2-\beta)
 &T(\beta,\alpha)&T(\alpha,\beta)\\
D(\alpha,\beta)&O(\alpha,\beta)&D(\pi/2-\alpha,\pi/2-\beta)
 &C_{\mathrm{w}}(\alpha,\beta)&D(\beta,\alpha)\\
O(\alpha,\beta)&D(\alpha,\beta)&O(\pi/2-\alpha,\pi/2-\beta)
 &C_{\mathrm{b}}(\alpha,\beta)&O(\alpha,\beta)\\
C_{\mathrm{w}}(\alpha,\beta)&C_{\mathrm{w}}(\pi/2-\alpha,\pi/2-\beta)&C_{\mathrm{b}}(\alpha,\beta)
 &D(\alpha,\beta)&C_{\mathrm{w}}(\beta,\alpha)\\
C_{\mathrm{b}}(\alpha,\beta)&C_{\mathrm{b}}(\pi/2-\alpha,\pi/2-\beta)&C_{\mathrm{w}}(\alpha,\beta)
 &O(\alpha,\beta)&C_{\mathrm{b}}(\beta,\alpha)
\end{array}
\end{equation}
for all $\alpha,\beta\in[0,\pi/2]$. Consequently, on $\mathcal T$ the
orbits are generated by the two complements
$\alpha\mapsto\pi/2-\alpha$, $\beta\mapsto\pi/2-\beta$ and by the
interchange $\alpha\leftrightarrow\beta$; every orbit containing a
sparse class meets $\mathcal D$, and two points of $\mathcal D$ lie in
the same orbit if and only if their angles are related by the group generated by the
simultaneous complement
$(\alpha,\beta)\mapsto(\pi/2-\alpha,\pi/2-\beta)$ and the interchange.
\end{proposition}

\subsection{The disc quotient and its two classical points}\label{sec:disc}
Choosing a representative of each orbit in $\mathcal T$ and in
$\mathcal D$ leads to two triangles, glued along one side. This is the content of the next theorem, illustrated in
Figure~\ref{fig:classical-disk}.

\begin{theorem}[The disc quotient]\label{thm:classical-disk}
The quotient $\mathcal W/G_{\mathrm{cl}}$ is homeomorphic to the closed
disc obtained from the triangles
\begin{equation}\label{eq:fundamental-triangles}
\Delta_{\mathcal T}=\{(\alpha,\beta):0\le\alpha\le\beta\le \pi/4\},\qquad
\Delta_{\mathcal D}=\{(\alpha,\beta):0\le\alpha\le\beta,\ \alpha+\beta\le \pi/2\}
\end{equation}
by the identification 
\begin{equation}\label{eq:quotient-gluing}
T(0,\beta)=D(\beta,\beta),\qquad \beta\in[0,\pi/4].
\end{equation}
Exactly two points are classical: $A$, represented by
$T(0,0)=D(0,0)$, and $B$, represented by $D(0,\pi/2)$. They are the
two essential equivalence classes of classical Sudoku.
\end{theorem}

\begin{figure}[tbp]
\centering
\begin{tikzpicture}[x=.95cm,y=.95cm,font=\small,
  t1/.style={circle,fill=black,inner sep=0pt,minimum size=5pt},
  t2/.style={circle,draw=black,fill=lightgray,line width=.7pt,inner sep=0pt,minimum size=5pt}]
% ---- sparse triangle, vertical side AB
\fill[orange!28] (0,0)--(0,4)--(2,2)--cycle;
\draw[thick,orange!75!black] (0,0)--(0,4)--(2,2);
\draw[very thick,violet] (0,0)--(2,2);
\draw[thin] (1.84,1.84)--(1.68,2)--(1.84,2.16);
\node[t1] at (0,0) {};
\node[t2] at (0,4) {};
\node[below left] at (0,0) {$A$};
\node[above left] at (0,4) {$B$};
\node[right] at (2,2) {$C$};
\node at (.72,2) {$\Delta_{\mathcal D}$};
\draw[->,gray!70] (-1.1,-.9)--(-.3,-.9) node[right,font=\scriptsize]{$\alpha$};
\draw[->,gray!70] (-1.1,-.9)--(-1.1,-.1) node[above,font=\scriptsize]{$\beta$};
% ---- tensor triangle, vertical side AC
\fill[blue!18] (5,0)--(5,2)--(7,2)--cycle;
\draw[thick,blue!65!black] (5,0)--(7,2)--(5,2);
\draw[very thick,violet] (5,0)--(5,2);
\draw[thin] (5,1.8)--(5.2,1.8)--(5.2,2);
\node[t1] at (5,0) {};
\node[below left] at (5,0) {$A$};
\node[above left] at (5,2) {$C$};
\node[above right] at (7,2) {$D$};
\node at (5.62,1.32) {$\Delta_{\mathcal T}$};
\draw[->,dashed,violet,thick] (1.3,.9)--(4.7,.9)
  node[midway,above,font=\scriptsize]{$D(\beta,\beta)=T(0,\beta)$};
\end{tikzpicture}

\smallskip
{\small
\begin{tabular}{@{}lcc@{}}
\toprule
Vertex & Coordinates in $\Delta_{\mathcal T}$
 & Coordinates in $\Delta_{\mathcal D}$\\
\midrule
$A$ & $(0,0)$ & $(0,0)$\\
$B$ & --- & $(0,\pi/2)$\\
$C$ & $(0,\pi/4)$ & $(\pi/4,\pi/4)$\\
$D$ & $(\pi/4,\pi/4)$ & ---\\
\bottomrule
\end{tabular}}
\caption{Diagram for the quotient of $\mathcal{W}$ by $G_{\mathrm{cl}}$. The
sparse triangle $\Delta_{\mathcal D}$ (orange) and the tensor triangle
$\Delta_{\mathcal T}$ (blue) are drawn at the same scale, with the
first angle as horizontal coordinate. The arrow indicates the
identification of the two violet sides $AC$ according
to~\eqref{eq:quotient-gluing}. The vertices $A$ (black circle) and
$B$ (grey circle) represent the classical classes of Type~1 and
Type~2.}
\label{fig:classical-disk}
\end{figure}

Representatives of the classical classes at $A$ and $B$ are
\[
\begin{array}{c@{\qquad\qquad}c}
A:&B:\\[2pt]
\left[\begin{array}{cc|cc}
1&2&4&3\\3&4&2&1\\\hline4&3&1&2\\2&1&3&4
\end{array}\right]&
\left[\begin{array}{cc|cc}
1&2&3&4\\3&4&2&1\\\hline4&3&1&2\\2&1&4&3
\end{array}\right],
\end{array}
\]
with symbol $i$ standing for $E_i$; they are the Type~1 and Type~2
$4\times 4$ Sudoku of~\cite{AFLT}. Both are limits of classes of cardinality
eight, but only $A$, which lies in the closed tensor region, is a
limit of classes of cardinality sixteen. The vertices $C$ and $D$ have
cardinality eight and sixteen, and every point other than $A$ and $B$
is nonclassical.

The two types can be recognised on the grid itself. In a classical
SudoQ the rows of the fourth block carry the same two pairs of
symbols as the rows of the first block, or its columns carry the same
pairs as the columns of the first block: indeed, in normalised form
the permutation matrix $W$ has the support pattern of $\mathcal D$ or $\mathcal O$ in
the first case, and of $\mathcal C_{\mathrm w}$ or $\mathcal C_{\mathrm b}$
in the second. The grid is of Type~1 if both hold, and of Type~2 if
only one does. In $A$ the fourth block repeats the first; in $B$ its
rows carry the pairs $\{1,2\}$ and $\{3,4\}$ of the first block, but
its columns carry $\{1,4\}$ and $\{2,3\}$ instead of $\{1,3\}$ and
$\{2,4\}$. By~\eqref{eq:sparse-intersections}, Type~1 grids are the
four corners of the tensor square, where $W$ is a tensor product of
two $2\times2$ permutation matrices, and Type~2 grids are the
remaining eight sparse corners (Table~\ref{tab:classes}). They are
marked by black and grey circles in Figures~\ref{fig:five-squares},
\ref{fig:cardinality} and~\ref{fig:classical-disk}.

\section{Fractional Sudoku of size two}\label{sec:realparam}
We now consider the outcome probabilities of the measurement of  a grid in the basis of its first block. This turns
a SudoQ into a fractional Sudoku whose first block is classical.

\begin{definition}\label{def:canonical-fs}
The \emph{fractional Sudoku} $\mathbf p$ of a SudoQ $\mathbf P$ of size two is
\begin{equation}\label{eq:canonical-fs}
p_{r,c}\bigl(2(a-1)+b\bigr)=\tr(P_{a,b}P_{r,c}),\qquad a,b\in\{1,2\},\ r,c \in\{1,2,3,4\}.
\end{equation}
\end{definition}

This is~\eqref{eq:image} with $f_{2(a-1)+b}\in\Ran P_{a,b}$. The array
$\mathbf p$ is invariant under unitary equivalence, and its first block
is classical, $p_{a,b}=e_{2(a-1)+b}$. Its matrix representation follows
directly from that of the normalised SudoQ. For a unitary matrix
$M=(m_{ij})_{i,j=1}^4$, write $|M|^2=(|m_{ij}|^2)_{i,j=1}^4$, where the squared modulus
is taken entrywise. Arranging the four probability vectors of each
block as columns, in the same cell order as before, gives
\[
\mathbf P\longleftrightarrow
\begin{array}{c|c}I_4&U\\\hline V&W\end{array}
\qquad\longmapsto\qquad
\mathbf p\longleftrightarrow
\begin{array}{c|c}I_4&|U|^2\\\hline |V|^2&|W|^2\end{array}.
\]
In each block matrix, rows index the measurement outcomes and columns
index the cells. Explicitly, for $r,c\in\{1,2\}$ and $k=1,\ldots,4$,
\begin{equation}\label{eq:fs-from-UVW}
p_{r,2+c}(k)=|u_{k,2(r-1)+c}|^2,\quad
p_{2+r,c}(k)=|v_{k,2(r-1)+c}|^2,\quad
p_{2+r,2+c}(k)=|w_{k,2(r-1)+c}|^2.
\end{equation}
The four block matrices are therefore unistochastic: by definition,
each is obtained by taking the squared moduli of the entries of a
unitary matrix. In particular, they are doubly stochastic, since
unitarity gives
\[
\sum_{i=1}^{4} (|M|^2)_{ij}=\sum_{i=1}^{4}|m_{ij}|^2=1,\qquad
\sum_{j=1}^{4} (|M|^2)_{ij}=\sum_{j=1}^{4}|m_{ij}|^2=1.
\]
The column sums express normalisation of the probability distribution
in each cell, and the row sums express the block constraint for each
outcome. The same argument applied to the orthonormal bases in the
rows and columns of the SudoQ proves the corresponding unistochasticity
statements of Lemma~\ref{lem:unistochastic}.

By Theorem~\ref{th:cardinality}, the cardinality of $\mathbf P$ is the
number of nonzero entries in the fourth block of $\mathbf p$.

\begin{theorem}[Fractional Sudoku of a SudoQ]\label{thm:fs}
Let $\mathbf P$, $\mathbf P'$ be SudoQ of size two, with fractional
Sudoku $\mathbf p$, $\mathbf p'$.
\begin{enumerate}
\item $\mathbf P\equiv\mathbf P'$ if and only if $\mathbf p=\mathbf p'$.
\item After replacing $\mathbf P$ by a unitarily equivalent normalised
grid, which leaves $\mathbf p$ unchanged, choose its matrix
representation as
\[
\mathbf P\longleftrightarrow
\begin{array}{c|c}
I_4&U\\\hline
V&W=X(\alpha,\beta)
\end{array},
\qquad \alpha,\beta\in[0,\pi/2],
\quad X\in\{T,D,O,C_{\mathrm w},C_{\mathrm b}\},
\]
as provided by Theorem~\ref{setW}, with $U,V$ given
by~\eqref{eq:angular-blocks}. Then, with $x=\cos^2(\alpha)$ and
$y=\cos^2(\beta)$,
\begin{equation}\label{eq:bilinear}
\mathbf p=xy\,S^X_{00}+x(1-y)\,S^X_{01}+(1-x)y\,S^X_{10}+(1-x)(1-y)\,S^X_{11},
\end{equation}
where, for $i,j\in\{0,1\}$, $S^X_{ij}$ denotes the classical Sudoku
obtained at the corner $X(i\pi/2,j\pi/2)$ of the family $X$, with
first block $\left[\begin{smallmatrix}1&2\\3&4\end{smallmatrix}\right]$,
as listed in Table~\ref{tab:classes}. Here each symbol $a$ is
identified with the probability vector $e_a$, so $S^X_{ij}$ is viewed
as a fractional Sudoku. In particular,
$\mathbf p$ is a convex combination of at most four classical Sudoku.
\end{enumerate}
\end{theorem}

Part~(1) says that the outcome probabilities of a measurement in the
basis of the first block determine the SudoQ up to unitary
equivalence. For the tensor family,
\eqref{eq:bilinear} has a transparent form: the fractional Sudoku of
$T(\alpha,\beta)$ is the grid of Example~\ref{ex:product} with $Q_a$, $L_a$ replaced by
$e_a\in\R^2$, with $\widetilde Q_1,\widetilde Q_2$ replaced by
$(x,1-x),(1-x,x)$ and $\widetilde L_1,\widetilde L_2$ by
$(y,1-y),(1-y,y)$, the tensor product being taken in
$\R^2\otimes\R^2=\R^4$, and $e_{2(a-1)+b}=e_a\otimes e_b$. The cells of the
second and third blocks thus carry two symbols, with weights $y,1-y$
and $x,1-x$, and those of the fourth block carry product distributions.

Not every fractional Sudoku with a classical first block arises from a SudoQ.

\begin{example}\label{ex:not-quantum}
Let $0<\lambda<1$. In $\lambda\,S^T_{00}+(1-\lambda)\,S^T_{11}$ the cells
$(3,1)$ and $(3,3)$ carry $\lambda e_4+(1-\lambda)e_2$ and
$\lambda e_1+(1-\lambda)e_4$, so $4$ is the only symbol to which both give
positive probability; in $\lambda\,S^T_{01}+(1-\lambda)\,S^T_{10}$ the same
cells carry $\lambda e_4+(1-\lambda)e_2$ and $\lambda e_2+(1-\lambda)e_3$,
and the only such symbol is $2$. By
Lemma~\ref{lem:unistochastic}, neither array arises from a SudoQ, in
any basis.
\end{example}

Let $\mathcal F$ be the set of fractional Sudoku of size two with first
block $\left[\begin{smallmatrix}e_1&e_2\\e_3&e_4\end{smallmatrix}\right]$.
It contains the fractional Sudoku of every SudoQ of size two.

\begin{proposition}\label{prop:polytope}
$\mathcal F$ is a polytope of dimension seven, whose vertices are the
twelve classical Sudoku of Table~\ref{tab:classes}. In particular, every
fractional Sudoku with a classical first block is a convex combination
of classical Sudoku.
\end{proposition}
\begin{remark}[Geometry]\label{rem:geometry}
By Theorem~\ref{thm:fs}, the map sending a SudoQ to its fractional
Sudoku embeds $\mathcal W$ in $\mathcal F$, the twelve classical classes
going to the vertices. The image is easier to describe than to draw.
\begin{enumerate}
\item For a sparse family $\mathcal X$ every entry of the fractional
Sudoku depends on one angle only, so~\eqref{eq:bilinear} is affine in
$(x,y)$, and $\mathcal X$ is mapped onto the parallelogram with vertices
$S^X_{00},S^X_{10},S^X_{11},S^X_{01}$. It is the two-dimensional face
of $\mathcal F$ on which the fourth block is supported in the pattern
of $\mathcal X$ in~\eqref{eq:support-patterns}.
These support restrictions set nonnegative probability
coordinates equal to zero and therefore define a face
of $\mathcal F$.
\item The four Type~1 vertices span a tetrahedron, and the tensor
family is mapped into it, onto the 
surface~\eqref{eq:bilinear}. Its
boundary consists of the edges $S^T_{00}S^T_{10}$, $S^T_{10}S^T_{11}$,
$S^T_{11}S^T_{01}$ and $S^T_{01}S^T_{00}$ of the tetrahedron, which are
the diagonals of the four parallelograms through their Type~1 vertices.
\item The other two edges of the tetrahedron, $S^T_{00}S^T_{11}$ and
$S^T_{01}S^T_{10}$, contain no fractional Sudoku of a SudoQ other than
their endpoints, by Example~\ref{ex:not-quantum}.
\end{enumerate}
The five-square model of Figure~\ref{fig:five-squares} thus sits in the
seven-dimensional polytope $\mathcal F$ as a two-dimensional saddle (the tensor family),
with four flat faces (the sparse families) hinged on its boundary.
\end{remark}

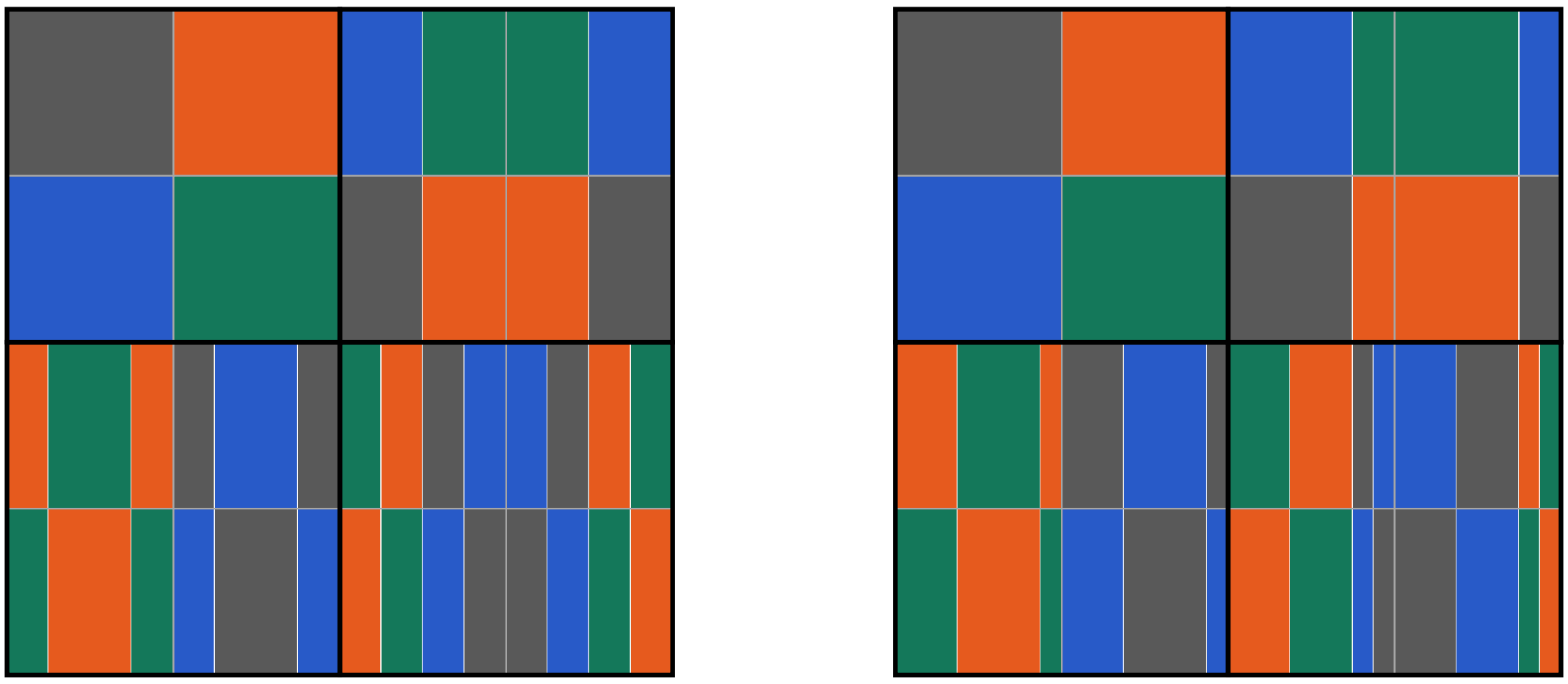
\begin{figure}[htbp]
\centering
\begin{minipage}{0.48\textwidth}
\centering
\begin{tikzpicture}[x=1cm,y=1cm]
\fill[ck0] (0.0000,4.5000) rectangle (1.5000,6.0000);
\fill[ck1] (1.5000,4.5000) rectangle (3.0000,6.0000);
\fill[ck2] (3.0000,4.5000) rectangle (3.7500,6.0000);
\draw[white,line width=0.6pt] (3.7500,4.5000) -- (3.7500,6.0000);
\fill[ck3] (3.7500,4.5000) rectangle (4.5000,6.0000);
\fill[ck3] (4.5000,4.5000) rectangle (5.2500,6.0000);
\draw[white,line width=0.6pt] (5.2500,4.5000) -- (5.2500,6.0000);
\fill[ck2] (5.2500,4.5000) rectangle (6.0000,6.0000);
\fill[ck2] (0.0000,3.0000) rectangle (1.5000,4.5000);
\fill[ck3] (1.5000,3.0000) rectangle (3.0000,4.5000);
\fill[ck0] (3.0000,3.0000) rectangle (3.7500,4.5000);
\draw[white,line width=0.6pt] (3.7500,3.0000) -- (3.7500,4.5000);
\fill[ck1] (3.7500,3.0000) rectangle (4.5000,4.5000);
\fill[ck1] (4.5000,3.0000) rectangle (5.2500,4.5000);
\draw[white,line width=0.6pt] (5.2500,3.0000) -- (5.2500,4.5000);
\fill[ck0] (5.2500,3.0000) rectangle (6.0000,4.5000);
\fill[ck1] (0.0000,1.5000) rectangle (0.3750,3.0000);
\draw[white,line width=0.6pt] (0.3750,1.5000) -- (0.3750,3.0000);
\fill[ck3] (0.3750,1.5000) rectangle (1.1250,3.0000);
\draw[white,line width=0.6pt] (1.1250,1.5000) -- (1.1250,3.0000);
\fill[ck1] (1.1250,1.5000) rectangle (1.5000,3.0000);
\fill[ck0] (1.5000,1.5000) rectangle (1.8750,3.0000);
\draw[white,line width=0.6pt] (1.8750,1.5000) -- (1.8750,3.0000);
\fill[ck2] (1.8750,1.5000) rectangle (2.6250,3.0000);
\draw[white,line width=0.6pt] (2.6250,1.5000) -- (2.6250,3.0000);
\fill[ck0] (2.6250,1.5000) rectangle (3.0000,3.0000);
\fill[ck3] (3.0000,1.5000) rectangle (3.3750,3.0000);
\draw[white,line width=0.6pt] (3.3750,1.5000) -- (3.3750,3.0000);
\fill[ck1] (3.3750,1.5000) rectangle (3.7500,3.0000);
\draw[white,line width=0.6pt] (3.7500,1.5000) -- (3.7500,3.0000);
\fill[ck0] (3.7500,1.5000) rectangle (4.1250,3.0000);
\draw[white,line width=0.6pt] (4.1250,1.5000) -- (4.1250,3.0000);
\fill[ck2] (4.1250,1.5000) rectangle (4.5000,3.0000);
\fill[ck2] (4.5000,1.5000) rectangle (4.8750,3.0000);
\draw[white,line width=0.6pt] (4.8750,1.5000) -- (4.8750,3.0000);
\fill[ck0] (4.8750,1.5000) rectangle (5.2500,3.0000);
\draw[white,line width=0.6pt] (5.2500,1.5000) -- (5.2500,3.0000);
\fill[ck1] (5.2500,1.5000) rectangle (5.6250,3.0000);
\draw[white,line width=0.6pt] (5.6250,1.5000) -- (5.6250,3.0000);
\fill[ck3] (5.6250,1.5000) rectangle (6.0000,3.0000);
\fill[ck3] (0.0000,0.0000) rectangle (0.3750,1.5000);
\draw[white,line width=0.6pt] (0.3750,0.0000) -- (0.3750,1.5000);
\fill[ck1] (0.3750,0.0000) rectangle (1.1250,1.5000);
\draw[white,line width=0.6pt] (1.1250,0.0000) -- (1.1250,1.5000);
\fill[ck3] (1.1250,0.0000) rectangle (1.5000,1.5000);
\fill[ck2] (1.5000,0.0000) rectangle (1.8750,1.5000);
\draw[white,line width=0.6pt] (1.8750,0.0000) -- (1.8750,1.5000);
\fill[ck0] (1.8750,0.0000) rectangle (2.6250,1.5000);
\draw[white,line width=0.6pt] (2.6250,0.0000) -- (2.6250,1.5000);
\fill[ck2] (2.6250,0.0000) rectangle (3.0000,1.5000);
\fill[ck1] (3.0000,0.0000) rectangle (3.3750,1.5000);
\draw[white,line width=0.6pt] (3.3750,0.0000) -- (3.3750,1.5000);
\fill[ck3] (3.3750,0.0000) rectangle (3.7500,1.5000);
\draw[white,line width=0.6pt] (3.7500,0.0000) -- (3.7500,1.5000);
\fill[ck2] (3.7500,0.0000) rectangle (4.1250,1.5000);
\draw[white,line width=0.6pt] (4.1250,0.0000) -- (4.1250,1.5000);
\fill[ck0] (4.1250,0.0000) rectangle (4.5000,1.5000);
\fill[ck0] (4.5000,0.0000) rectangle (4.8750,1.5000);
\draw[white,line width=0.6pt] (4.8750,0.0000) -- (4.8750,1.5000);
\fill[ck2] (4.8750,0.0000) rectangle (5.2500,1.5000);
\draw[white,line width=0.6pt] (5.2500,0.0000) -- (5.2500,1.5000);
\fill[ck3] (5.2500,0.0000) rectangle (5.6250,1.5000);
\draw[white,line width=0.6pt] (5.6250,0.0000) -- (5.6250,1.5000);
\fill[ck1] (5.6250,0.0000) rectangle (6.0000,1.5000);
\draw[gray!70,thin] (0.0000,4.5000) rectangle (1.5000,6.0000);
\draw[gray!70,thin] (1.5000,4.5000) rectangle (3.0000,6.0000);
\draw[gray!70,thin] (3.0000,4.5000) rectangle (4.5000,6.0000);
\draw[gray!70,thin] (4.5000,4.5000) rectangle (6.0000,6.0000);
\draw[gray!70,thin] (0.0000,3.0000) rectangle (1.5000,4.5000);
\draw[gray!70,thin] (1.5000,3.0000) rectangle (3.0000,4.5000);
\draw[gray!70,thin] (3.0000,3.0000) rectangle (4.5000,4.5000);
\draw[gray!70,thin] (4.5000,3.0000) rectangle (6.0000,4.5000);
\draw[gray!70,thin] (0.0000,1.5000) rectangle (1.5000,3.0000);
\draw[gray!70,thin] (1.5000,1.5000) rectangle (3.0000,3.0000);
\draw[gray!70,thin] (3.0000,1.5000) rectangle (4.5000,3.0000);
\draw[gray!70,thin] (4.5000,1.5000) rectangle (6.0000,3.0000);
\draw[gray!70,thin] (0.0000,0.0000) rectangle (1.5000,1.5000);
\draw[gray!70,thin] (1.5000,0.0000) rectangle (3.0000,1.5000);
\draw[gray!70,thin] (3.0000,0.0000) rectangle (4.5000,1.5000);
\draw[gray!70,thin] (4.5000,0.0000) rectangle (6.0000,1.5000);
\draw[very thick] (0.0000,3.0000) rectangle (3.0000,6.0000);
\draw[very thick] (3.0000,3.0000) rectangle (6.0000,6.0000);
\draw[very thick] (0.0000,0.0000) rectangle (3.0000,3.0000);
\draw[very thick] (3.0000,0.0000) rectangle (6.0000,3.0000);
\end{tikzpicture}
\par\smallskip\textbf{(a)}
\end{minipage}\hfill
\begin{minipage}{0.48\textwidth}
\centering
\begin{tikzpicture}[x=1cm,y=1cm]
\fill[ck0] (0.0000,4.5000) rectangle (1.5000,6.0000);
\fill[ck1] (1.5000,4.5000) rectangle (3.0000,6.0000);
\fill[ck2] (3.0000,4.5000) rectangle (4.1250,6.0000);
\draw[white,line width=0.6pt] (4.1250,4.5000) -- (4.1250,6.0000);
\fill[ck3] (4.1250,4.5000) rectangle (4.5000,6.0000);
\fill[ck3] (4.5000,4.5000) rectangle (5.6250,6.0000);
\draw[white,line width=0.6pt] (5.6250,4.5000) -- (5.6250,6.0000);
\fill[ck2] (5.6250,4.5000) rectangle (6.0000,6.0000);
\fill[ck2] (0.0000,3.0000) rectangle (1.5000,4.5000);
\fill[ck3] (1.5000,3.0000) rectangle (3.0000,4.5000);
\fill[ck0] (3.0000,3.0000) rectangle (4.1250,4.5000);
\draw[white,line width=0.6pt] (4.1250,3.0000) -- (4.1250,4.5000);
\fill[ck1] (4.1250,3.0000) rectangle (4.5000,4.5000);
\fill[ck1] (4.5000,3.0000) rectangle (5.6250,4.5000);
\draw[white,line width=0.6pt] (5.6250,3.0000) -- (5.6250,4.5000);
\fill[ck0] (5.6250,3.0000) rectangle (6.0000,4.5000);
\fill[ck1] (0.0000,1.5000) rectangle (0.5625,3.0000);
\draw[white,line width=0.6pt] (0.5625,1.5000) -- (0.5625,3.0000);
\fill[ck3] (0.5625,1.5000) rectangle (1.3125,3.0000);
\draw[white,line width=0.6pt] (1.3125,1.5000) -- (1.3125,3.0000);
\fill[ck1] (1.3125,1.5000) rectangle (1.5000,3.0000);
\fill[ck0] (1.5000,1.5000) rectangle (2.0625,3.0000);
\draw[white,line width=0.6pt] (2.0625,1.5000) -- (2.0625,3.0000);
\fill[ck2] (2.0625,1.5000) rectangle (2.8125,3.0000);
\draw[white,line width=0.6pt] (2.8125,1.5000) -- (2.8125,3.0000);
\fill[ck0] (2.8125,1.5000) rectangle (3.0000,3.0000);
\fill[ck3] (3.0000,1.5000) rectangle (3.5625,3.0000);
\draw[white,line width=0.6pt] (3.5625,1.5000) -- (3.5625,3.0000);
\fill[ck1] (3.5625,1.5000) rectangle (4.1250,3.0000);
\draw[white,line width=0.6pt] (4.1250,1.5000) -- (4.1250,3.0000);
\fill[ck0] (4.1250,1.5000) rectangle (4.3125,3.0000);
\draw[white,line width=0.6pt] (4.3125,1.5000) -- (4.3125,3.0000);
\fill[ck2] (4.3125,1.5000) rectangle (4.5000,3.0000);
\fill[ck2] (4.5000,1.5000) rectangle (5.0625,3.0000);
\draw[white,line width=0.6pt] (5.0625,1.5000) -- (5.0625,3.0000);
\fill[ck0] (5.0625,1.5000) rectangle (5.6250,3.0000);
\draw[white,line width=0.6pt] (5.6250,1.5000) -- (5.6250,3.0000);
\fill[ck1] (5.6250,1.5000) rectangle (5.8125,3.0000);
\draw[white,line width=0.6pt] (5.8125,1.5000) -- (5.8125,3.0000);
\fill[ck3] (5.8125,1.5000) rectangle (6.0000,3.0000);
\fill[ck3] (0.0000,0.0000) rectangle (0.5625,1.5000);
\draw[white,line width=0.6pt] (0.5625,0.0000) -- (0.5625,1.5000);
\fill[ck1] (0.5625,0.0000) rectangle (1.3125,1.5000);
\draw[white,line width=0.6pt] (1.3125,0.0000) -- (1.3125,1.5000);
\fill[ck3] (1.3125,0.0000) rectangle (1.5000,1.5000);
\fill[ck2] (1.5000,0.0000) rectangle (2.0625,1.5000);
\draw[white,line width=0.6pt] (2.0625,0.0000) -- (2.0625,1.5000);
\fill[ck0] (2.0625,0.0000) rectangle (2.8125,1.5000);
\draw[white,line width=0.6pt] (2.8125,0.0000) -- (2.8125,1.5000);
\fill[ck2] (2.8125,0.0000) rectangle (3.0000,1.5000);
\fill[ck1] (3.0000,0.0000) rectangle (3.5625,1.5000);
\draw[white,line width=0.6pt] (3.5625,0.0000) -- (3.5625,1.5000);
\fill[ck3] (3.5625,0.0000) rectangle (4.1250,1.5000);
\draw[white,line width=0.6pt] (4.1250,0.0000) -- (4.1250,1.5000);
\fill[ck2] (4.1250,0.0000) rectangle (4.3125,1.5000);
\draw[white,line width=0.6pt] (4.3125,0.0000) -- (4.3125,1.5000);
\fill[ck0] (4.3125,0.0000) rectangle (4.5000,1.5000);
\fill[ck0] (4.5000,0.0000) rectangle (5.0625,1.5000);
\draw[white,line width=0.6pt] (5.0625,0.0000) -- (5.0625,1.5000);
\fill[ck2] (5.0625,0.0000) rectangle (5.6250,1.5000);
\draw[white,line width=0.6pt] (5.6250,0.0000) -- (5.6250,1.5000);
\fill[ck3] (5.6250,0.0000) rectangle (5.8125,1.5000);
\draw[white,line width=0.6pt] (5.8125,0.0000) -- (5.8125,1.5000);
\fill[ck1] (5.8125,0.0000) rectangle (6.0000,1.5000);
\draw[gray!70,thin] (0.0000,4.5000) rectangle (1.5000,6.0000);
\draw[gray!70,thin] (1.5000,4.5000) rectangle (3.0000,6.0000);
\draw[gray!70,thin] (3.0000,4.5000) rectangle (4.5000,6.0000);
\draw[gray!70,thin] (4.5000,4.5000) rectangle (6.0000,6.0000);
\draw[gray!70,thin] (0.0000,3.0000) rectangle (1.5000,4.5000);
\draw[gray!70,thin] (1.5000,3.0000) rectangle (3.0000,4.5000);
\draw[gray!70,thin] (3.0000,3.0000) rectangle (4.5000,4.5000);
\draw[gray!70,thin] (4.5000,3.0000) rectangle (6.0000,4.5000);
\draw[gray!70,thin] (0.0000,1.5000) rectangle (1.5000,3.0000);
\draw[gray!70,thin] (1.5000,1.5000) rectangle (3.0000,3.0000);
\draw[gray!70,thin] (3.0000,1.5000) rectangle (4.5000,3.0000);
\draw[gray!70,thin] (4.5000,1.5000) rectangle (6.0000,3.0000);
\draw[gray!70,thin] (0.0000,0.0000) rectangle (1.5000,1.5000);
\draw[gray!70,thin] (1.5000,0.0000) rectangle (3.0000,1.5000);
\draw[gray!70,thin] (3.0000,0.0000) rectangle (4.5000,1.5000);
\draw[gray!70,thin] (4.5000,0.0000) rectangle (6.0000,1.5000);
\draw[very thick] (0.0000,3.0000) rectangle (3.0000,6.0000);
\draw[very thick] (3.0000,3.0000) rectangle (6.0000,6.0000);
\draw[very thick] (0.0000,0.0000) rectangle (3.0000,3.0000);
\draw[very thick] (3.0000,0.0000) rectangle (6.0000,3.0000);
\end{tikzpicture}
\par\smallskip\textbf{(b)}
\end{minipage}
\caption{Fractional Sudoku of (a) $T(\pi/4,\pi/4)$ and
(b) $T(\pi/4,\pi/3)$. Each cell is divided into regions of areas
$p_{r,c}(a)$, $a=1,\ldots,4$, one colour per symbol, arranged in four vertical bands
showing $S^T_{11}$, $S^T_{01}$, $S^T_{00}$, $S^T_{10}$, of widths
$\frac14,\frac14,\frac14,\frac14$ in~(a) and
$\frac38,\frac38,\frac18,\frac18$ in~(b).}
\label{fig:ex-pair}
\end{figure}

\emph{Visualisation of fractional Sudoku.} Following~\cite{HZ2026}, we draw a fractional Sudoku
by dividing every cell into regions of areas $p_{r,c}(a)$, the symbols
$1,2,3,4$ being drawn in grey, orange, blue and green. By~\eqref{eq:bilinear},
the fractional Sudoku of $T(\alpha,\beta)$ splits into four vertical
bands of widths $(1-x)(1-y)$, $x(1-y)$, $xy$, $(1-x)y$, showing
$S^T_{11}$, $S^T_{01}$, $S^T_{00}$, $S^T_{10}$. Each band being a
Sudoku, every colour covers a cell exactly once when the cells of a
row, column or block are superposed. This order of the bands minimises
the number of rectangles, forty in both examples of
Figure~\ref{fig:ex-pair}. The grid $T(\pi/4,\pi/4)$ of~\cite{KZ}, made
of four basis states, eight equal-weight superpositions of two and four
equal-weight superpositions of four basis states, has $x=y=\frac12$:
the cells of the first block are whole, those of the second and third
halved and those of the fourth quartered, as in Fig.~1c
of~\cite{HZ2026}. For $T(\pi/4,\pi/3)$, $x=\frac12$, $y=\frac14$ and
\[
\mathbf p=\tfrac18\,S^T_{00}+\tfrac38\,S^T_{01}+\tfrac18\,S^T_{10}+\tfrac38\,S^T_{11}.
\]

\FloatBarrier

\section{Proofs}\label{sec:proofs}
In this section we abbreviate, for a SudoQ of size two, the row and
column sums of the first and of the fourth block:
\begin{equation}\label{eq:rank-two-sums}
\begin{aligned}
F_a&=P_{a,1}+P_{a,2},& G_b&=P_{1,b}+P_{2,b},\\
B_r&=P_{2+r,3}+P_{2+r,4},& C_c&=P_{3,2+c}+P_{4,2+c},
\end{aligned}\qquad a,b,r,c\in\{1,2\}.
\end{equation}
These are rank-two projections with
$F_1+F_2=G_1+G_2=B_1+B_2=C_1+C_2=I_4$, since the entries of a block
are mutually orthogonal. In this notation Theorem~\ref{thm:WUV} states
that a completion exists if and only if $F_1C_1$ and $G_1B_1$ are
rank-one projections, and that then $P_{r,2+c}=F_r^\perp C_c^\perp$
and $P_{2+r,c}=B_r^\perp G_c^\perp$.

\subsection{Minimal cardinality and reconstruction}

\begin{proof}[Proof of Theorem~\ref{prop:commuting-classical}]
$(1)\Rightarrow(2)$. The $N^2$ projections of the first row are
distinct, so if $c(\mathbf P)=N^2$ every entry is one of them.
Conjugating by a unitary that maps $\Ran P_{1,j}$ onto $\C e_j$ for
every $j$ gives a classical SudoQ.

$(2)\Rightarrow(3)$. Canonical projections commute, and commutation is
preserved by simultaneous conjugation.

$(3)\Rightarrow(1)$. If $P_{r,c}$ commutes with $P_{1,j}$, then
$P_{r,c}P_{1,j}$ is the projection onto
$\Ran P_{r,c}\cap\Ran P_{1,j}$, hence either $0$ or
$P_{r,c}=P_{1,j}$. As $\sum_{j=1}^{N^2}P_{r,c}P_{1,j}=P_{r,c}\neq0$, every entry
equals some $P_{1,j}$.
\end{proof}

\begin{proof}[Proof of Theorem~\ref{thm:WUV}]
The product of two projections is a projection if and only if they
commute (if $QR$ is self-adjoint, $QR=(QR)^*=RQ$), and its rank is
then its trace; this gives the equivalence of the two formulations.

\emph{Necessity and uniqueness.} In a completion, row $r$ and column
$2+c$ give $\sum_{j=1}^{2}P_{r,2+j}=F_r^\perp$ and
$\sum_{i=1}^{2}P_{i,2+c}=C_c^\perp$. The four projections of the second block
being mutually orthogonal, the product of these two sums is
\[
F_r^\perp C_c^\perp=C_c^\perp F_r^\perp=P_{r,2+c},
\qquad\text{and likewise}\qquad
B_r^\perp G_c^\perp=G_c^\perp B_r^\perp=P_{2+r,c}.
\]
This is~\eqref{eq:projector-reconstruction}, so the completion is
unique; for $r=c=2$ it gives $F_1C_1=P_{2,4}$ and $B_1G_1=P_{4,2}$.

\emph{Sufficiency.} Let $Q,R$ be commuting rank-two projections with
$\tr(QR)=1$. Then $QR$, $QR^\perp$, $Q^\perp R$, $Q^\perp R^\perp$ are
mutually orthogonal projections summing to $I_4$, each of trace one.
For $(Q,R)=(F_1,C_1)$, whose complements are $F_2,C_2$, these are
the four entries~\eqref{eq:projector-reconstruction} of the second
block; for $(G_1,B_1)$ they are those of the third. Finally
$\sum_{j=1}^{2}P_{r,2+j}=F_r^\perp$, $\sum_{i=1}^{2}P_{i,2+c}=C_c^\perp$,
$\sum_{j=1}^{2}P_{2+r,j}=B_r^\perp$ and $\sum_{i=1}^{2}P_{2+i,c}=G_c^\perp$, so every
row and every column sums to $I_4$.
\end{proof}

\subsection{Cardinality and the overlap invariant}
We use two elementary facts.

\begin{lemma}\label{lem:rank-one}
Let $P,Q,R$ be rank-one projections with $PQ=0$ and $(P+Q)R=R$. Then
\begin{align}
&\tr(PR)+\tr(QR)=1,\label{eq:rank-one-overlap-sum}\\
&\|PRQ\|^2=\tr(PR)\tr(QR).\label{eq:rank-one-offdiagonal-norm}
\end{align}
In particular,
\begin{equation}\label{eq:rank-one-zero-overlaps}
\tr(PR)=0\Rightarrow R=Q,\qquad \tr(QR)=0\Rightarrow R=P,
\end{equation}
\begin{equation}\label{eq:rank-one-positive-overlaps}
PRQ\neq0\iff 0<\tr(PR)<1\iff0<\tr(QR)<1.
\end{equation}
\end{lemma}

\begin{proof}
Write $R=vv^*$; then $v=Pv+Qv$ and $PRQ=(Pv)(Qv)^*$.
\end{proof}

\begin{lemma}\label{lem:overlap-factorisation}
Let $\mathbf P$ be a SudoQ of size two. For all $a,b,r,c\in\{1,2\}$,
\begin{equation}\label{eq:overlap-product}
\tr(P_{a,b}P_{2+r,2+c})=\tr(P_{a,b}B_r)\,\tr(P_{a,b}C_c),
\end{equation}
\begin{equation}\label{eq:projector-propagation}
P_{a,b}B_1P_{3-a,b}C_1P_{3-a,3-b}
=P_{a,b}C_1P_{a,3-b}B_1P_{3-a,3-b}.
\end{equation}
\end{lemma}

\begin{proof}
By Theorem~\ref{thm:WUV} and its proof, each $B_r$ commutes with each
$G_j$ and each $C_c$ with each $F_i$. Since $P_{a,b}\le F_a,G_b$, this
gives
\begin{equation}\label{eq:block-selection}
P_{a,b}B_rP_{i,j}=0\ \ (j\neq b),\qquad
P_{i,j}C_cP_{a,b}=0\ \ (i\neq a).
\end{equation}
Write $P_{2+r,2+c}=B_rC_c$, insert $I_4=\sum_{i,j=1}^{2}P_{i,j}$ between
$B_r$ and $C_c$, and use~\eqref{eq:block-selection}: only $(i,j)=(a,b)$
survives, and $P_{a,b}XP_{a,b}=\tr(P_{a,b}X)P_{a,b}$
gives~\eqref{eq:overlap-product}. Inserting the same resolution of the
identity in $P_{a,b}B_1C_1P_{3-a,3-b}=P_{a,b}C_1B_1P_{3-a,3-b}$
gives~\eqref{eq:projector-propagation}.
\end{proof}

\begin{proof}[Proof of Theorem~\ref{th:cardinality}]
For a rank-one projection $Q$ on $\C^4$ put
\begin{equation}\label{eq:projector-support}
\supp(Q)=\{(a,b)\in\{1,2\}^2:\tr(P_{a,b}Q)>0\},\qquad
s(Q)=\#\supp(Q).
\end{equation}
Since $\sum_{a,b=1}^{2}\tr(P_{a,b}Q)=1$, we have $s(Q)=1$ if and only if
$Q$ is one of the $P_{a,b}$. Both quantities are invariant under
simultaneous unitary conjugation.
We will prove that 
$$
c(\mathbf P)=\sum_{r,c=1}^{2} s(P_{2+r,2+c})
=\#\{(a,b,r,c)\in\{1,2\}^4:\tr(P_{a,b}P_{2+r,2+c})>0\},
$$
and exactly one of the following holds:
\begin{enumerate}
\item all four $s(P_{2+r,2+c})$ equal one, and $c(\mathbf P)=4$;
\item two of them equal one and two equal two, and $c(\mathbf P)=6$;
\item all four equal two, and $c(\mathbf P)=8$;
\item all four equal four, and $c(\mathbf P)=16$.
\end{enumerate}
Recall that $B_rC_c=P_{2+r,2+c}$ and, by
Theorem~\ref{thm:WUV}, $F_aC_1=P_{3-a,4}$, $G_bB_1=P_{4,3-b}$ are
rank-one. Put
\[
x_a=P_{a,1}C_1P_{a,2},\qquad y_b=P_{1,b}B_1P_{2,b},\qquad a,b\in\{1,2\}.
\]
Lemma~\ref{lem:rank-one} with $(P,Q,R)=(P_{a,1},P_{a,2},F_aC_1)$ and
$(P_{1,b},P_{2,b},G_bB_1)$ shows that
\begin{equation}\label{eq:test-dichotomy}
\begin{aligned}
x_a&=0\iff F_aC_1\in\{P_{a,1},P_{a,2}\},&
x_a&\neq0\iff 0<\tr(P_{a,b}C_1)<1\ \ (b=1,2),\\
y_b&=0\iff G_bB_1\in\{P_{1,b},P_{2,b}\},&
y_b&\neq0\iff 0<\tr(P_{a,b}B_1)<1\ \ (a=1,2).
\end{aligned}
\end{equation}

\emph{Case 1: $x_{a'}\neq0$ and $y_{b'}\neq0$ for some $a',b'$.}
In~\eqref{eq:projector-propagation} with $a=3-a'$, $b=b'$, the
left-hand side is the product of $P_{3-a',b'}B_1P_{a',b'}$ and
$P_{a',b'}C_1P_{a',3-b'}$, which are $y_{b'}$ and $x_{a'}$ up to
adjoints, through the rank-one $P_{a',b'}$; it is therefore nonzero.
Hence so are both factors on the right-hand side, that is $x_{3-a'}$
and $y_{3-b'}$ up to adjoints. Thus all $x_a,y_b$ are nonzero, and
by~\eqref{eq:test-dichotomy}, $C_2=C_1^\perp$, $B_2=B_1^\perp$
and~\eqref{eq:overlap-product}, all sixteen overlaps
$\tr(P_{a,b}P_{2+r,2+c})$ are positive. Moreover,
by~\eqref{eq:projector-reconstruction},
$\tr(P_{a,b}P_{r,2+c})=\tr(P_{a,b}F_{3-r}C_{3-c})$ vanishes unless
$a=3-r$ and is then positive, and similarly for the third block, so
\[
\begin{aligned}
\supp(P_{r,c})&=\{(r,c)\},&
\supp(P_{r,2+c})&=\{3-r\}\times\{1,2\},\\
\supp(P_{2+r,c})&=\{1,2\}\times\{3-c\},&
\supp(P_{2+r,2+c})&=\{1,2\}^2.
\end{aligned}
\]
Equal projections have equal supports, so entries of different blocks
are distinct; entries of the same block are orthogonal. Hence
$c(\mathbf P)=16=\sum_{r,c=1}^{2}s(P_{2+r,2+c})$.

\emph{Case 2: $x_{a'}\neq0$ for some $a'$, and $y_1=y_2=0$.} By
\eqref{eq:test-dichotomy}, $B_1=G_1B_1+G_2B_1$ is one of
$F_1,F_2,P_{1,1}+P_{2,2},P_{2,1}+P_{1,2}$. For the last two,
$P_{a,1}[B_1,C_1]P_{a,2}=\pm x_a$, which is nonzero for $a=a'$,
contradicting $[B_1,C_1]=0$. Hence $\{B_1,B_2\}=\{F_1,F_2\}$, and
by~\eqref{eq:projector-reconstruction} every entry
$F_{3-r}C_{3-c}$ of the second block is an entry $B_iC_j$ of the
fourth, and every entry $B_{3-r}G_{3-c}$ of the third is an entry
$F_iG_j=P_{i,j}$ of the first. The fourth block consists of the two
pairs $\{F_aC_1,F_aC_2\}$, $a=1,2$, with $F_aC_1+F_aC_2=F_a$, and these
pairs are disjoint since $F_1F_2=0$. By~\eqref{eq:test-dichotomy}, if
$x_a=0$ the pair is $\{P_{a,1},P_{a,2}\}$, with supports of size one;
if $x_a\neq0$ both members have support $\{a\}\times\{1,2\}$, so they
are two new projections with supports of size two. Therefore
$c(\mathbf P)=6=1+1+2+2$ if exactly one $x_a$ is nonzero, and
$c(\mathbf P)=8=2+2+2+2$ if both are.

\emph{Case 3: $x_1=x_2=0$, and $y_{b'}\neq0$ for some $b'$.}
Transposition of the grid exchanges $F_a\leftrightarrow G_a$ and
$B_r\leftrightarrow C_r$, hence $x\leftrightarrow y$, and preserves
both sides of~\eqref{eq:intrinsic-cardinality}; we are back in Case 2.

\emph{Case 4: all $x_a,y_b$ vanish.} By~\eqref{eq:test-dichotomy},
$C_1=\sum_{a=1}^{2}F_aC_1$ and $B_1=\sum_{b=1}^{2}G_bB_1$ are sums of first-block
projections, and so are $C_2,B_2$. Hence every $B_rC_c$, and
by~\eqref{eq:projector-reconstruction} every entry of the grid, is a
first-block projection, and $c(\mathbf P)=4=\sum_{r,c=1}^{2}s(P_{2+r,2+c})$.

The four cases are exhaustive and mutually exclusive. In each case
the counting identity~\eqref{eq:intrinsic-cardinality} holds, and the
only possible cardinalities are $4,6,8,16$.
\end{proof}

\begin{proof}[Proof of Theorem~\ref{cor:complete-overlaps}]
Overlaps are invariant under unitary conjugation. Conversely, let
$\mathbf P$, $\mathbf P'$ have the same overlaps. After conjugating
$\mathbf P'$ we may assume $P'_{a,b}=P_{a,b}$ for all $a,b\in\{1,2\}$; this
changes neither its overlaps nor its class. Write $B=B_1$, $C=C_1$, and
$B'$, $C'$ for the same sums in $\mathbf P'$. It suffices to find a
unitary $\Gamma$ commuting with every $P_{a,b}$ such that
\begin{equation}\label{eq:BC-match}
\Gamma B\Gamma^*=B',\qquad \Gamma C\Gamma^*=C'.
\end{equation}
Indeed, \eqref{eq:BC-match} gives $\Gamma B_r\Gamma^*=B'_r$ and $\Gamma C_c\Gamma^*=C'_c$,
hence $\Gamma P_{2+r,2+c}\Gamma^*=\Gamma B_rC_c\Gamma^*=P'_{2+r,2+c}$; since $\Gamma$ fixes
$F_r$ and $G_c$, \eqref{eq:projector-reconstruction} gives the same
for the other two blocks.

Every operator $M$ satisfies $M=\sum_{a,b,i,j=1}^{2}P_{a,b}MP_{i,j}$. For
$M=B,B'$ only the terms with $j=b$ survive, and for $M=C,C'$ only
those with $i=a$, by~\eqref{eq:block-selection}. The diagonal terms
agree: $P_{a,b}BP_{a,b}=\tr(P_{a,b}B)P_{a,b}$ and
$\tr(P_{a,b}B)=\sum_{c=1}^{2}\tr(P_{a,b}P_{3,2+c})$ is an overlap sum, and
similarly for $C$. The off-diagonal terms are $P_{1,b}BP_{2,b}$,
$P_{a,1}CP_{a,2}$ and their adjoints; by
Lemma~\ref{lem:rank-one}, applied as in~\eqref{eq:test-dichotomy},
their norms are $\tr(P_{1,b}B)^{1/2}\tr(P_{2,b}B)^{1/2}$ and
$\tr(P_{a,1}C)^{1/2}\tr(P_{a,2}C)^{1/2}$, again determined by the
overlaps. Being maps between rank-one ranges, they differ from their
primed counterparts by phases:
\[
P_{a,1}C'P_{a,2}=\xi_a\,P_{a,1}CP_{a,2},\qquad
P_{1,b}B'P_{2,b}=\eta_b\,P_{1,b}BP_{2,b},\qquad|\xi_a|=|\eta_b|=1,
\]
with $\xi_a=1$ or $\eta_b=1$ when the products vanish.

Take $\Gamma=\sum_{a,b=1}^{2}\lambda_{a,b}P_{a,b}$ with $|\lambda_{a,b}|=1$. It is
unitary, commutes with the $P_{a,b}$, and satisfies
$P_{a,b}\Gamma M\Gamma^*P_{i,j}=\lambda_{a,b}\overline{\lambda_{i,j}}P_{a,b}MP_{i,j}$ for every operator $M$.
Hence~\eqref{eq:BC-match} holds as soon as
\begin{equation}\label{eq:phase-equations}
\lambda_{a,1}\overline{\lambda_{a,2}}=\xi_a
\ \text{ if } P_{a,1}CP_{a,2}\neq0,\qquad
\lambda_{1,b}\overline{\lambda_{2,b}}=\eta_b
\ \text{ if } P_{1,b}BP_{2,b}\neq0.
\end{equation}
By Case 1 in the proof of Theorem~\ref{th:cardinality}, either all four
products are nonzero, or both $P_{a,1}CP_{a,2}$ vanish, or both
$P_{1,b}BP_{2,b}$ vanish. In the last two cases~\eqref{eq:phase-equations}
is solved by $\lambda_{1,b}=1$, $\lambda_{2,b}=\overline{\eta_b}$,
respectively $\lambda_{a,1}=1$, $\lambda_{a,2}=\overline{\xi_a}$. In
the first case, \eqref{eq:projector-propagation} with $a=b=1$ for both
grids reads
\[
P_{1,1}BP_{2,1}CP_{2,2}=P_{1,1}CP_{1,2}BP_{2,2}\neq0,\qquad
\eta_1\xi_2\,P_{1,1}BP_{2,1}CP_{2,2}=\xi_1\eta_2\,P_{1,1}CP_{1,2}BP_{2,2},
\]
so $\eta_1\xi_2=\xi_1\eta_2$, and
$\lambda_{1,1}=1$, $\lambda_{1,2}=\overline{\xi_1}$,
$\lambda_{2,1}=\overline{\eta_1}$, $\lambda_{2,2}=\overline{\eta_1\xi_2}$
solve all four equations.
\end{proof}

\begin{proof}[Proof of Corollary~\ref{cor:real}]
 Normalise the first block to the canonical projections. The grids $\mathbf P$ and $\overline{\mathbf P}$ have identical opposite-block overlaps, so Theorem~\ref{cor:complete-overlaps} gives
\[
\overline{P_{r,c}}=\Gamma P_{r,c}\Gamma^*.
\]
Since both first blocks are canonical, $\Gamma$ is diagonal. Choose a diagonal unitary $D$ with $D^2=\Gamma$. Then
\[
\overline{DP_{r,c}D^*}
=D^*\overline{P_{r,c}}D
=D^*\Gamma P_{r,c}\Gamma^*D
=DP_{r,c}D^*.
\]
Hence every transformed projection is real.  
\end{proof}

\subsection{Matrix representation and the parameter space}

\begin{proof}[Proof of Proposition~\ref{prop:normalization}]
Conjugation by $R^*$ normalises the first block. A unitary relating two
normalised grids fixes every $E_j$, hence is a diagonal $\Delta$, and the
right factors account for the choice of representatives. Conversely,
right factors change no projection, and the common left factor
conjugates every entry by $\Delta$ while fixing the first block.
\end{proof}

\begin{proof}[Proof of Theorem~\ref{cor:matrix-reconstruction}]
For the normalised first block, $F_r=E_{2r-1}+E_{2r}$ and
$G_c=E_c+E_{c+2}$, while $B_r=WF_rW^*$ and $C_c=WG_cW^*$. In the
bases $(e_{2r-1},e_{2r})$ of $\Ran F_r$ and $(e_r,e_{r+2})$ of
$\Ran G_r$, the compressions $F_rC_cF_r$ and $G_rB_cG_r$ have matrices
$W_{[r,c]}{W_{[r,c]}}^*$ and $W^{[r,c]}{W^{[r,c]}}^*$. If a
completion exists, $F_rC_c$ and $G_rB_c$ are rank-one projections by
Theorem~\ref{thm:WUV}, and the determinants vanish.

Conversely, assume they vanish. By unitarity,
$W_{[r,1]}{W_{[r,1]}}^*+W_{[r,2]}{W_{[r,2]}}^*=I_2$ is a sum
of two positive matrices of rank at most one, hence of two
complementary rank-one projections (neither vanishes, and a unit
vector in the kernel of one is fixed by the other); likewise for
$W^{[r,1]}$ and $W^{[r,2]}$. So $F_1C_1F_1$ is a projection of trace one. If $QRQ$ is a
projection, then $[Q,R]=0$, because
$QRQ-(QRQ)^2=(Q^\perp RQ)^*(Q^\perp RQ)$. Hence $F_1$, $C_1$ commute
with $\tr(F_1C_1)=1$, and likewise $G_1$, $B_1$;
Theorem~\ref{thm:WUV} gives a unique completion.

Under $W\mapsto D_LWD_R$ the minors are multiplied by phases, so
admissibility is a property of $[W]_\sim$. The right factor changes no
projection, and the left one fixes the first block and conjugates the
fourth by $D_L$; by uniqueness of the completion, the two grids are
conjugate by $D_L$. Together with Proposition~\ref{prop:normalization}
this shows that $[W]_\sim$ is a complete invariant of the class, and
every class is attained since every SudoQ can be normalised.
\end{proof}

\begin{proof}[Proof of Theorem~\ref{setW}]
Direct substitution shows that each of the five families satisfies
the defining conditions of $\mathcal W$.
It remains to prove that every element of $\mathcal W$ belongs
to one of these families.

\emph{Full matrices.} Let $W\in\mathcal W$ be full, i.e. all the entries of $W$ are nonzero. Rephase its first
row to $(a,b,c,d)$ with positive entries and its first column to
$(a,-e,-f,g)^{\mathsf T}$ with $e,f,g>0$. With $x=w_{22}$, $y=w_{33}$,
the eight minors of~\eqref{systemW} force
\[
W=\begin{pmatrix}
a&b&c&d\\
-e&x&-ce/a&dx/b\\
-f&-bf/a&y&dy/c\\
g&-gx/e&-gy/f&adgxy/(bcef)
\end{pmatrix}.
\]
Orthogonality of the first row to the second and third gives
\[
x=\frac{be(a^2+c^2)}{a(b^2+d^2)}>0,\qquad
y=\frac{cf(a^2+b^2)}{a(c^2+d^2)}>0,
\]
and orthogonality of the second and third rows gives
\[
0=\frac{ef(a^2d^2-b^2c^2)(a^2+b^2+c^2+d^2)}
{a^2(b^2+d^2)(c^2+d^2)},
\]
so $ad=bc$. The trace conditions $\tr(F_1C_1)=\tr(G_1B_1)=1$ read
\[
(a^2+c^2)(1+e^2/a^2)=1,\qquad
(a^2+b^2)(1+f^2/a^2)=1,
\]
and with $a^2+b^2+c^2+d^2=1$ and $ad=bc$ they give $e=b$, $f=c$, and
then $g=d$. Hence $x=y=a$ and
\[
W=\begin{pmatrix}
a&b&c&d\\-b&a&-d&c\\-c&-d&a&b\\d&-c&-b&a
\end{pmatrix}=A(\alpha)\otimes A(\beta),
\qquad \tan\alpha=c/a,\quad\tan\beta=b/a,
\]
with $\alpha,\beta\in(0,\pi/2)$.

\emph{Matrices with a zero.} Each matrix of a sparse family
satisfies~\eqref{systemW}, since every selected $2\times2$ submatrix
has a zero row or column. The selected minors use either consecutive
rows and equal-parity columns, or the reverse. Hence permuting rows,
or columns, by $(1\,2)(3\,4)$ and $(1\,3)(2\,4)$, transposing, and
applying $\mathscr S$ preserve unitarity and~\eqref{systemW}, and
permute the four sparse families. Use them to move a zero to
$w_{11}$. The first equation gives $w_{13}w_{21}=0$; if $w_{21}=0$,
replace $W$ by $\mathscr S(W^{\mathsf T})$. So the first row is
$(0,b,0,d)$. If $b,d\neq0$, the second and fifth equations give
$w_{31}=w_{33}=0$, so rows $1,3$ span $\operatorname{span}\{e_2,e_4\}$
and rows $2,4$ its complement: this is $\mathcal C_{\mathrm b}$. If
$bd=0$, move the nonzero entry of the first row to position $(1,4)$
and make it $1$; the rest of column $4$ vanishes, and the third and
fifth equations give $w_{22}=w_{33}=0$:
\[
\begin{pmatrix}
0&0&0&1\\ a&0&b&0\\ c&d&0&0\\ e&f&g&0
\end{pmatrix}.
\]
Orthogonality of rows $2,3$ gives $a\overline c=0$. If $a\neq0$, then
$c=0$, $|d|=1$, $f=0$, which is $\mathcal C_{\mathrm b}$; if $a=0$,
then $|b|=1$, $g=0$, which is $\mathcal O$. Undoing the symmetries,
every solution with a zero lies in a sparse family.

\emph{Intersections.} A full matrix lies in no sparse family. In
$T(\alpha,\beta)$ the off-diagonal $2\times2$ blocks vanish if and only
if $\sin(\alpha)=0$, the diagonal ones if and only if $\cos(\alpha)=0$,
the black checkerboard entries if and only if $\sin(\beta)=0$, and the
white ones if and only if $\cos(\beta)=0$. At $\alpha=0$,
$T=\diag(A(\beta),A(\beta))=D(\beta,\beta)$; at $\alpha=\pi/2$ the
nonzero blocks are $A(\beta)$ and $-A(\beta)$, and a diagonal phase
gives $O(\beta,\beta)$; applying $\mathscr S$, which exchanges the
tensor factors, gives the other two identities
in~\eqref{eq:four-gluings}. Since diagonal factors preserve zeros, a
class common to two sparse families has support in both patterns
of~\eqref{eq:support-patterns}. The patterns of $\mathcal D$ and
$\mathcal O$ are disjoint, as are those of $\mathcal C_{\mathrm w}$ and
$\mathcal C_{\mathrm b}$; for the other four pairs the common positions
form a permutation pattern, whose unique class is the corresponding
corner of $\mathcal T$. This proves~\eqref{eq:sparse-intersections};
since any three sparse families contain a disjoint pair, the
statements on triple and higher intersections follow. The same
exchange of the two types of minors gives
$\mathscr S(\mathcal W)=\mathcal W$.

\emph{Uniqueness of the angles.} For $T(\alpha,\beta)$, boundary
included, $\cos^2(\alpha)=|w_{11}|^2+|w_{12}|^2$ and
$\cos^2(\beta)=|w_{11}|^2+|w_{31}|^2$ are invariant under diagonal
factors. For a sparse family, each nonzero $2\times2$ block (after
undoing $\mathscr S$) is reduced separately, and the modulus of its
upper-left entry fixes its angle.

\emph{Topology.} Each map $(\alpha,\beta)\mapsto X(\alpha,\beta)$ is
continuous and injective, the five maps cover $\mathcal W$, and their
coincidences are generated by~\eqref{eq:four-gluings}. They induce a
continuous bijection from the compact quotient of five squares onto
$\mathcal W$, which is Hausdorff as a subspace of the orbit space of
the compact group $\mathbb T_4\times\mathbb T_4$ acting on $\U(4)$. Any continuous
bijection from a compact space to a Hausdorff space is a
homeomorphism~\cite[Thm.~26.6]{Munkres}.
\end{proof}

\subsection{Cell symmetries}

\begin{proof}[Proof of Proposition~\ref{lem:cell-action}]
Let $Z=\diag(1,-1)$. For $\alpha\in[0,\pi/2]$,
\begin{equation}\label{eq:angle-action-identities}
A(\alpha)\Pi=Z A(\pi/2-\alpha),\qquad
A(\alpha)^{\mathsf T}=Z A(\alpha)Z,\qquad
JA(\alpha)=A(\alpha)J,
\end{equation}
and diagonal signs do not change a class. Substituting the
representatives~\eqref{eq:angle-representatives}
into~\eqref{eq:generator-fourth-blocks} gives the columns $r$, $c$,
$\tau$; for instance
$T(\alpha,\beta)(\Pi\otimes I_2)=(Z\otimes I_2)T(\pi/2-\alpha,\beta)$,
and conjugation by $K$ exchanges the tensor factors and maps $D,O$ to
$C_{\mathrm w},C_{\mathrm b}$. For $b$, \eqref{eq:angular-blocks}
together with $J^*=-J$, $J^2=-I_2$ gives
\[
\renewcommand{\arraystretch}{1.5}
\begin{array}{c|c}
W&V^*U\\\hline
T(\alpha,\beta)&A(\alpha)^{\mathsf T}\otimes A(\beta)\\
D(\alpha,\beta)&\diag(-A(\beta),A(\alpha))\\
O(\alpha,\beta)&O(\alpha,\beta)\\
C_{\mathrm w}(\alpha,\beta)
 &K\diag(-A(\beta)^{\mathsf T},A(\alpha)^{\mathsf T})K\\
C_{\mathrm b}(\alpha,\beta)
 &K\begin{pmatrix}0&A(\beta)^{\mathsf T}\\A(\alpha)^{\mathsf T}&0\end{pmatrix}K
\end{array}
\]
and removing signs by~\eqref{eq:angle-action-identities} yields
$T(\alpha,\beta)$, $D(\beta,\alpha)$, $O(\alpha,\beta)$,
$C_{\mathrm w}(\beta,\alpha)$, $C_{\mathrm b}(\beta,\alpha)$. These
identities hold on the closed squares.

On $\mathcal T$ every generator complements one angle, interchanges
them, or acts trivially, and $r,c,\tau$ realise all three. On the
sparse families every entry of the table changes $(\alpha,\beta)$ by
the simultaneous complement, the interchange, or not at all; these
commute and are involutions. Since $r\cdot O(\alpha,\beta)$,
$\tau\cdot C_{\mathrm w}(\alpha,\beta)$ and
$r\cdot(\tau\cdot C_{\mathrm b}(\alpha,\beta))$ all equal $D(\alpha,\beta)$,
every sparse orbit meets $\mathcal D$, and by uniqueness of the angles
(Theorem~\ref{setW}) it meets $\mathcal D$ exactly as stated.
Changing description at an intersection adds nothing: outside
$\mathcal T$ a class lies in a single sparse family
by~\eqref{eq:sparse-intersections}, and the tensor moves send
$D(\beta,\beta)=T(0,\beta)$ to some $T(0,\gamma)$ only for
$\gamma\in\{\beta,\pi/2-\beta\}$, which is the simultaneous complement
on the diagonal of $\mathcal D$.
\end{proof}

\begin{proof}[Proof of Theorem~\ref{thm:classical-disk}]
Complementing each angle into $[0,\pi/4]$ and ordering them gives a
unique representative of every $\mathcal T$-orbit in
$\Delta_{\mathcal T}$; ordering the angles and, if $\alpha+\beta>\pi/2$,
complementing both and reordering gives a unique representative of
every $\mathcal D$-orbit in $\Delta_{\mathcal D}$. Boundary points are
fixed by the corresponding reflections, so uniqueness persists there.
By Theorem~\ref{setW} and Proposition~\ref{lem:cell-action}, the only
orbits meeting both triangles are those of~\eqref{eq:quotient-gluing},
which glue a full side of one triangle to a full side of the other.
The result is a closed disc, and the continuous bijection onto
$\mathcal W/G_{\mathrm{cl}}$ is a homeomorphism, since the domain is
compact and the quotient of a Hausdorff space by a finite group is
Hausdorff. Every sparse corner lies in the orbit of a corner of
$\mathcal D$, and these form two orbits under the simultaneous
complement, represented by $(0,0)$ and $(0,\pi/2)$. Accordingly, the
twelve classical points of Section~\ref{sec:matrix-cardinality} split
into an orbit of size four, the corners of $\mathcal T$, and one of
size eight.
\end{proof}

\subsection{Fractional Sudoku}

\begin{proof}[Proof of Lemma~\ref{lem:unistochastic}]
Write $P_{r,c}=u_{r,c}u_{r,c}^*$. The vectors $u_{r,1},\ldots,u_{r,N^2}$ form an orthonormal
basis, so $(\langle f_a,u_{r,c}\rangle)_{c,a}$ is a unitary matrix, and $p_{r,c}(a)$ are the
squares of the moduli of its entries. If $a$ is the only symbol with
$p_{r,c}(a)\,p_{r,c'}(a)>0$, all other terms of
$\langle u_{r,c},u_{r,c'}\rangle=\sum_{b=1}^{N^2}\overline{\langle f_b,u_{r,c}\rangle}
\langle f_b,u_{r,c'}\rangle$ vanish, and $\langle u_{r,c},u_{r,c'}\rangle
=\overline{\langle f_a,u_{r,c}\rangle}\langle f_a,u_{r,c'}\rangle\neq0$, contradicting
orthogonality. Columns and blocks are treated in the same way.
\end{proof}

\begin{proof}[Proof of Theorem~\ref{thm:fs}]
(1) The fractional Sudoku $\mathbf p$ of $\mathbf P$ is a unitary invariant, and its fourth block consists of
the overlaps $\tr(P_{a,b}P_{2+r,2+c})$, which form a complete invariant by
Theorem~\ref{cor:complete-overlaps}.

(2) By Theorem~\ref{setW} and~\eqref{eq:fs-from-UVW}, $\mathbf p$ consists of the
squared moduli of the entries of $U$, $V$, $W$ in~\eqref{eq:angular-blocks}. These matrices
are built from $A(\alpha)$, $A(\beta)$ and signed permutation matrices by products, tensor
products and block arrangements, so every entry of $\mathbf p$ has the form
$f(\alpha)g(\beta)$ with $f\in\{0,1,\cos^2(\alpha),\sin^2(\alpha)\}$ and
$g\in\{0,1,\cos^2(\beta),\sin^2(\beta)\}$. As a function of $(x,y)\in[0,1]^2$ it is therefore
affine in each variable, and coincides with the bilinear interpolation of its values at
the four corners. At a corner, $A(\alpha)$ and $A(\beta)$, hence $U$, $V$, $W$, are signed
permutation matrices, the grid is classical, and $\mathbf p$ is the classical
Sudoku $S^X_{ij}$.
\end{proof}

\begin{proof}[Proof of Proposition~\ref{prop:polytope}]
The first block forces the second and third blocks to have the form
\begin{equation}\label{eq:blocks23}
\left[\begin{array}{cc}
te_3+(1-t)e_4&(1-t)e_3+te_4\\
ue_1+(1-u)e_2&(1-u)e_1+ue_2
\end{array}\right],\qquad
\left[\begin{array}{cc}
ve_2+(1-v)e_4&we_1+(1-w)e_3\\
(1-v)e_2+ve_4&(1-w)e_1+we_3
\end{array}\right],
\end{equation}
with $t,u,v,w\in[0,1]$.
For each symbol $a\in\{1,2,3,4\}$, put
\[
m_a=p_{3,3}(a),\qquad
R_a=1-p_{3,1}(a)-p_{3,2}(a),\qquad
C_a=1-p_{1,3}(a)-p_{2,3}(a).
\]
The four entries for symbol $a$ in the fourth block, in row
order, are
\[
m_a,\quad R_a-m_a,\quad C_a-m_a,\quad 1-R_a-C_a+m_a.
\]
Thus their nonnegativity is equivalent to
\[
\max\{0,R_a+C_a-1\}\leq m_a\leq\min\{R_a,C_a\},
\qquad a=1,\ldots,4.
\]
In addition, $\sum_{a=1}^{4}m_a=1$; the other three
cell-normalisation conditions follow from
$\sum_{a=1}^{4}R_a=\sum_{a=1}^{4}C_a=2$.
These conditions describe $\mathcal F$ as a bounded intersection
of finitely many affine half-spaces, hence as a polytope.
There are $4+3=7$ affine parameters, and all the inequalities
are strict at
\[
t=u=v=w=\frac12,\qquad m_a=\frac14
\quad(a=1,\ldots,4).
\]
Therefore $\dim\mathcal F=7$.

In a classical Sudoku, $t,u,v,w\in\{0,1\}$, and every symbol
must occupy the cell of the fourth block that avoids its row
in the third block and its column in the second.
A direct check shows that these four cells are distinct exactly
when $t=u$ or $v=w$, giving the twelve grids of
Table~\ref{tab:classes}.
Being points with coordinates zero and one of a polytope
contained in the unit cube, they are vertices.

It remains to write every $\mathbf p\in\mathcal F$ as a convex
combination of these classical Sudoku.
Put
\[
y=(y_1,y_2,y_3,y_4)=(t,1-v,1-u,w),\qquad
n=(n_1,n_2,n_3,n_4)=(m_4,m_2,m_1,m_3),
\]
with indices taken modulo four, and set
$(a_1,a_2,a_3,a_4)=(4,2,1,3)$.
A direct computation shows that, for each $i\in\{1,2,3,4\}$,
the four numbers
\[
n_i,\qquad b_i=y_i-n_i,\qquad
y_{i+1}-y_i+n_i,\qquad 1-y_{i+1}-n_i
\]
are, in an order depending on $i$, the four probabilities
$p_{2+r,2+c}(a_i)$, $r,c\in\{1,2\}$.
Hence they are nonnegative.
Moreover, $\sum_{i=1}^{4}n_i=1$, and $\mathbf p$ is an affine
function of $(y,n)$.

The twelve classical Sudoku are the points
\[
y=\mathbf 1_{\{k-\ell+1,\ldots,k\}},\qquad n=e_k,
\qquad k\in\{1,2,3,4\},\quad \ell\in\{1,2,3\},
\]
where $\mathbf 1_{\{k-\ell+1,\ldots,k\}}$ is the vector
whose entries are one at the $\ell$ consecutive positions
$k-\ell+1,\ldots,k$, with indices taken modulo four,
and zero elsewhere.
Indeed, these twelve pairs satisfy the constraints above
and give distinct zero-one probability arrays, so they
coincide with the twelve classical completions.
Call this Sudoku $S_{k,\ell}$.
It suffices to find weights $\mu_{k,\ell}\geq0$ such that
\[
\sum_{\ell=1}^{3}\mu_{k,\ell}=n_k,\qquad
\sum_{k=1}^{4}\sum_{\ell=1}^{3}
\mu_{k,\ell}\mathbf 1_{\{k-\ell+1,\ldots,k\}}=y,
\]
for then
\[
\mathbf p=
\sum_{k=1}^{4}\sum_{\ell=1}^{3}\mu_{k,\ell}S_{k,\ell}.
\]

Put
\[
H_k=\max\{0,b_{k-2}-n_{k-1}\},\qquad
G_k=b_{k-1}-H_{k+1},
\]
and define
\[
\mu_{k,3}=H_k,\qquad
\mu_{k,2}=G_k-H_k,\qquad
\mu_{k,1}=n_k-G_k.
\]
Then $\sum_{\ell=1}^{3}\mu_{k,\ell}=n_k$, and the total weight
of the intervals containing $j$ is
\[
n_j+G_{j+1}+H_{j+2}=n_j+b_j=y_j.
\]
Clearly $\mu_{k,3}\geq0$, while
\[
\mu_{k,1}=n_k-b_{k-1}+H_{k+1}\geq0
\]
by the definition of $H_{k+1}$.
Finally, $\mu_{k,2}\geq0$ is equivalent to
$H_k+H_{k+1}\leq b_{k-1}$.
Using the identity
\[
\max\{0,A\}+\max\{0,B\}=\max\{0,A,B,A+B\},
\]
this is equivalent to the four inequalities
\[
0\leq b_{k-1},\qquad
b_{k-1}-n_k\leq b_{k-1},\qquad
b_{k-2}-n_{k-1}\leq b_{k-1},\qquad
b_{k-2}\leq n_{k-1}+n_k.
\]
The first two follow from $b_i,n_i\geq0$.
The third is
\[
y_{k-1}-y_{k-2}+n_{k-2}\geq0,
\]
and, using $\sum_{i=1}^{4}n_i=1$, the fourth is
\[
1-y_{k+2}-n_{k+1}\geq0.
\]
Both hold by the nonnegativity of the fourth-block probabilities.
Thus the coefficients $\mu_{k,\ell}$ are nonnegative and sum
to one, proving the required convex decomposition.
\end{proof}

\section{Conclusions and open questions}\label{sec:conclusions}
We have given a complete description of the SudoQ of size two. Two opposite blocks determine a SudoQ
(Theorem~\ref{thm:WUV}), and the sixteen overlaps between them form a complete invariant,
whose support counts the distinct states (Theorems~\ref{th:cardinality}
and~\ref{cor:complete-overlaps}). The equivalence classes form the five-square model of
Theorem~\ref{setW}; every class has a real representative, and modulo the classical cell
symmetries one obtains a closed disc with two classical points
(Theorem~\ref{thm:classical-disk}). The measurement outcome probabilities of every cell in the basis of the first block (or any other block) 
give a fractional Sudoku, which determines the grid up to unitary
equivalence and is a convex combination of at most four classical Sudoku (Theorem~\ref{thm:fs}). Two natural questions remain open.
\begin{enumerate}
\item Do Theorems~\ref{thm:WUV}, \ref{cor:complete-overlaps} and~\ref{thm:fs} have analogues for
$N\geq3$?
\item Which fractional Sudoku arise from SudoQ? For $N=2$ and the basis of the first block
the answer is the image of the five-square model (Remark~\ref{rem:geometry}); in a general
basis Lemma~\ref{lem:unistochastic} gives a necessary condition, but we know no
characterisation.
\end{enumerate}
These and other questions are left for future work.

\appendix

%%%%%%%%%%%%%%%%%%%%%%%%%%%%%%%%%%%%%%%%%%%%%%%%%%%%%%%%%%%%%%%%%%%%%%%%%%%%%%%%
%% APPENDIX. Paste this block after Section 9 (Conclusions and open
%% questions) and before \section*{Statements and Declarations}.
%% It uses only \diag, the colours ck0--ck3 and the tikz library `patterns',
%% which are already in the preamble.
%%%%%%%%%%%%%%%%%%%%%%%%%%%%%%%%%%%%%%%%%%%%%%%%%%%%%%%%%%%%%%%%%%%%%%%%%%%%%%%%
\appendix
\section{A worked example}\label{app:example}
We include for the reader's convenience an example of SudoQ: we start from a vector representation, we provide the representation in terms of the matrices $I_4,U,V,W$ of
Section~\ref{sec:matrix-representation}, and a visualisation in terms of \emph{signed fractional Sudoku}. 
The SudoQ is the normal form $T(\alpha,\beta)$ of Theorem~\ref{setW} with
$\alpha=\pi/4$ and $\beta=\pi/3$, so that
$\cos(\alpha)=\sin(\alpha)=1/\sqrt2$, $\cos(\beta)=1/2$ and
$\sin(\beta)=\sqrt3/2$. Its unsigned pattern is
Figure~\ref{fig:ex-pair}(b).

\subsection{The grid of vectors}
Let $e_1,\ldots,e_4$ be the canonical basis of $\C^4$ and, as in the
introduction, name some superpositions,
\[
\xi_\pm=\frac{e_1\pm e_3}{\sqrt2},\qquad
\eta_\pm=\frac{e_2\pm e_4}{\sqrt2}.
\]
Consider the grid
\begin{equation}\label{eq:app-grid}
\begingroup
\renewcommand{\arraystretch}{1.4}
\setlength{\arraycolsep}{8pt}
\left[\begin{array}{cc|cc}
e_1&e_2&\frac{\sqrt3}2\,e_3+\frac12\,e_4&-\frac12\,e_3+\frac{\sqrt3}2\,e_4\\
e_3&e_4&-\frac{\sqrt3}2\,e_1-\frac12\,e_2&\frac12\,e_1-\frac{\sqrt3}2\,e_2\\\hline
\eta_+&-\xi_+&\frac12\,\xi_--\frac{\sqrt3}2\,\eta_-&\frac{\sqrt3}2\,\xi_-+\frac12\,\eta_-\\
-\eta_-&\xi_-&\frac12\,\xi_+-\frac{\sqrt3}2\,\eta_+&\frac{\sqrt3}2\,\xi_++\frac12\,\eta_+
\end{array}\right]
\endgroup
\end{equation}
with entries $u_{r,c}$. Every row, column and block is an orthonormal
basis, so $P_{r,c}=u_{r,c}u_{r,c}^*$ defines a SudoQ $\mathbf P$. It is real,
and its sixteen projections are distinct, so $c(\mathbf P)=16$. The signs in
\eqref{eq:app-grid} are those of the matrices below; only the projections
matter.

\subsection{The matrices \texorpdfstring{$I_4,U,V,W$}{I4, U, V, W}}
Arrange the vectors of each block as the columns of a matrix, in the order
$(r,c)=(1,1)$, $(1,2)$, $(2,1)$, $(2,2)$ of Definition~\ref{defn:fourmat}.
The first block gives $R=I_4$, and the others give exactly the matrices
of~\eqref{eq:angular-blocks}:
\[
\mathbf P\longleftrightarrow\begin{array}{c|c}I_4&U\\\hline V&W\end{array},
\qquad
U=J\otimes\bigl(A(\beta)J\bigr),\quad
V=\bigl(A(\alpha)J\bigr)\otimes J,\quad
W=A(\alpha)\otimes A(\beta).
\]
Explicitly,
{\small\begin{align*}
I_4&= \begin{pmatrix}1&0&0&0\\0&1&0&0\\0&0&1&0\\0&0&0&1\end{pmatrix}&U&=\frac12\begin{pmatrix}0&0&-\sqrt3&1\\0&0&-1&-\sqrt3\\\sqrt3&-1&0&0\\1&\sqrt3&0&0\end{pmatrix},\\
V&=\frac1{\sqrt2}\begin{pmatrix}0&-1&0&1\\1&0&-1&0\\0&-1&0&-1\\1&0&1&0\end{pmatrix},
&W&=\frac{\sqrt2}4\begin{pmatrix}1&\sqrt3&1&\sqrt3\\-\sqrt3&1&-\sqrt3&1\\-1&-\sqrt3&1&\sqrt3\\\sqrt3&-1&-\sqrt3&1\end{pmatrix}.
\end{align*}}
The fourth block consists of the products of the columns of $A(\alpha)$
and of $A(\beta)$, since $W=A(\alpha)\otimes A(\beta)$.

\subsection{The signed fractional Sudoku}
By Section~\ref{sec:realparam}, the fractional Sudoku of $\mathbf P$ has the
matrices $I_4$, $|U|^2$, $|V|^2$, $|W|^2$, where $|M|^2$ is the entrywise
squared modulus. Keeping the signs, we call
\[
\mathbf p_\mathrm{signed}=
\begin{array}{c|c}\begin{pmatrix}1&0&0&0\\0&1&0&0\\0&0&1&0\\0&0&0&1\end{pmatrix}&\frac14\begin{pmatrix}0&0&-3&1\\0&0&-1&-3\\3&-1&0&0\\1&3&0&0\end{pmatrix}\\\hline \frac12\begin{pmatrix}0&-1&0&1\\1&0&-1&0\\0&-1&0&-1\\1&0&1&0\end{pmatrix}&\frac18\begin{pmatrix}1&3&1&3\\-3&1&-3&1\\-1&-3&1&3\\3&-1&-3&1\end{pmatrix}\end{array}
\]
the \emph{signed fractional Sudoku} associated with the chosen real vector representatives of $\mathbf P$. 
In each matrix, row $k$ is the symbol $k$ and column $j$ is the $j$th cell of
the block; the entry is the signed probability of the symbol in that cell. The signed fractional Sudoku is represented in
 Figure~\ref{fig:app-signed}(a) where the field of colour $k$ in a cell, of area
$p_{r,c}(k)$, is hatched when the corresponding entry is negative. Removing
the hatching gives Figure~\ref{fig:ex-pair}(b).

\subsection{Diagonal unitaries on the left and on the right}
Let $\Gamma=\diag(1,-1,1,-1)$, and define the real normalised grids
\[
\mathbf P'\longleftrightarrow
\begin{array}{c|c}I_4&\Gamma U\\\hline\Gamma V&\Gamma W\end{array},
\qquad
\mathbf P''\longleftrightarrow
\begin{array}{c|c}I_4&\Gamma UD_U\\\hline\Gamma VD_V&\Gamma WD_W\end{array},
\]
with
\[
D_U=D_V=\diag(1,1,-1,1),\qquad D_W=\diag(1,-1,1,1).
\]
By Proposition~\ref{prop:normalization} they are unitarily equivalent to
$\mathbf P$; explicitly, $P'_{r,c}=\Gamma P_{r,c}\Gamma^*=P''_{r,c}$ for all
$r,c$. Thus $\mathbf P'$ and $\mathbf P''$ are the same SudoQ, the image of
$\mathbf P$ under $\Gamma$, but their vectors differ: $\mathbf P''$ is
obtained from $\mathbf P'$ by reversing the vectors in the cells $(2,3)$,
$(4,1)$ and $(3,4)$, namely the third column of $U$, the third column of $V$
and the second column of $W$.

\begin{figure}[!hb]
\centering
\begin{minipage}{0.325\textwidth}
\centering
\begin{tikzpicture}[x=0.8cm,y=0.8cm]
\fill[ck0] (0.0000,4.5000) rectangle (1.5000,6.0000);
\fill[ck1] (1.5000,4.5000) rectangle (3.0000,6.0000);
\fill[ck2] (3.0000,4.5000) rectangle (4.1250,6.0000);
\draw[white,line width=0.6pt] (4.1250,4.5000) -- (4.1250,6.0000);
\fill[ck3] (4.1250,4.5000) rectangle (4.5000,6.0000);
\fill[ck3] (4.5000,4.5000) rectangle (5.6250,6.0000);
\draw[white,line width=0.6pt] (5.6250,4.5000) -- (5.6250,6.0000);
\fill[ck2] (5.6250,4.5000) rectangle (6.0000,6.0000);
\fill[pattern=north east lines, pattern color=white] (5.6250,4.5000) rectangle (6.0000,6.0000);
\fill[ck2] (0.0000,3.0000) rectangle (1.5000,4.5000);
\fill[ck3] (1.5000,3.0000) rectangle (3.0000,4.5000);
\fill[ck0] (3.0000,3.0000) rectangle (4.1250,4.5000);
\fill[pattern=north east lines, pattern color=white] (3.0000,3.0000) rectangle (4.1250,4.5000);
\draw[white,line width=0.6pt] (4.1250,3.0000) -- (4.1250,4.5000);
\fill[ck1] (4.1250,3.0000) rectangle (4.5000,4.5000);
\fill[pattern=north east lines, pattern color=white] (4.1250,3.0000) rectangle (4.5000,4.5000);
\fill[ck1] (4.5000,3.0000) rectangle (5.6250,4.5000);
\fill[pattern=north east lines, pattern color=white] (4.5000,3.0000) rectangle (5.6250,4.5000);
\draw[white,line width=0.6pt] (5.6250,3.0000) -- (5.6250,4.5000);
\fill[ck0] (5.6250,3.0000) rectangle (6.0000,4.5000);
\fill[ck1] (0.0000,1.5000) rectangle (0.5625,3.0000);
\draw[white,line width=0.6pt] (0.5625,1.5000) -- (0.5625,3.0000);
\fill[ck3] (0.5625,1.5000) rectangle (1.3125,3.0000);
\draw[white,line width=0.6pt] (1.3125,1.5000) -- (1.3125,3.0000);
\fill[ck1] (1.3125,1.5000) rectangle (1.5000,3.0000);
\fill[ck0] (1.5000,1.5000) rectangle (2.0625,3.0000);
\fill[pattern=north east lines, pattern color=white] (1.5000,1.5000) rectangle (2.0625,3.0000);
\draw[white,line width=0.6pt] (2.0625,1.5000) -- (2.0625,3.0000);
\fill[ck2] (2.0625,1.5000) rectangle (2.8125,3.0000);
\fill[pattern=north east lines, pattern color=white] (2.0625,1.5000) rectangle (2.8125,3.0000);
\draw[white,line width=0.6pt] (2.8125,1.5000) -- (2.8125,3.0000);
\fill[ck0] (2.8125,1.5000) rectangle (3.0000,3.0000);
\fill[pattern=north east lines, pattern color=white] (2.8125,1.5000) rectangle (3.0000,3.0000);
\fill[ck3] (3.0000,1.5000) rectangle (3.5625,3.0000);
\draw[white,line width=0.6pt] (3.5625,1.5000) -- (3.5625,3.0000);
\fill[ck1] (3.5625,1.5000) rectangle (4.1250,3.0000);
\fill[pattern=north east lines, pattern color=white] (3.5625,1.5000) rectangle (4.1250,3.0000);
\draw[white,line width=0.6pt] (4.1250,1.5000) -- (4.1250,3.0000);
\fill[ck0] (4.1250,1.5000) rectangle (4.3125,3.0000);
\draw[white,line width=0.6pt] (4.3125,1.5000) -- (4.3125,3.0000);
\fill[ck2] (4.3125,1.5000) rectangle (4.5000,3.0000);
\fill[pattern=north east lines, pattern color=white] (4.3125,1.5000) rectangle (4.5000,3.0000);
\fill[ck2] (4.5000,1.5000) rectangle (5.0625,3.0000);
\fill[pattern=north east lines, pattern color=white] (4.5000,1.5000) rectangle (5.0625,3.0000);
\draw[white,line width=0.6pt] (5.0625,1.5000) -- (5.0625,3.0000);
\fill[ck0] (5.0625,1.5000) rectangle (5.6250,3.0000);
\draw[white,line width=0.6pt] (5.6250,1.5000) -- (5.6250,3.0000);
\fill[ck1] (5.6250,1.5000) rectangle (5.8125,3.0000);
\draw[white,line width=0.6pt] (5.8125,1.5000) -- (5.8125,3.0000);
\fill[ck3] (5.8125,1.5000) rectangle (6.0000,3.0000);
\fill[pattern=north east lines, pattern color=white] (5.8125,1.5000) rectangle (6.0000,3.0000);
\fill[ck3] (0.0000,0.0000) rectangle (0.5625,1.5000);
\draw[white,line width=0.6pt] (0.5625,0.0000) -- (0.5625,1.5000);
\fill[ck1] (0.5625,0.0000) rectangle (1.3125,1.5000);
\fill[pattern=north east lines, pattern color=white] (0.5625,0.0000) rectangle (1.3125,1.5000);
\draw[white,line width=0.6pt] (1.3125,0.0000) -- (1.3125,1.5000);
\fill[ck3] (1.3125,0.0000) rectangle (1.5000,1.5000);
\fill[ck2] (1.5000,0.0000) rectangle (2.0625,1.5000);
\fill[pattern=north east lines, pattern color=white] (1.5000,0.0000) rectangle (2.0625,1.5000);
\draw[white,line width=0.6pt] (2.0625,0.0000) -- (2.0625,1.5000);
\fill[ck0] (2.0625,0.0000) rectangle (2.8125,1.5000);
\draw[white,line width=0.6pt] (2.8125,0.0000) -- (2.8125,1.5000);
\fill[ck2] (2.8125,0.0000) rectangle (3.0000,1.5000);
\fill[pattern=north east lines, pattern color=white] (2.8125,0.0000) rectangle (3.0000,1.5000);
\fill[ck1] (3.0000,0.0000) rectangle (3.5625,1.5000);
\fill[pattern=north east lines, pattern color=white] (3.0000,0.0000) rectangle (3.5625,1.5000);
\draw[white,line width=0.6pt] (3.5625,0.0000) -- (3.5625,1.5000);
\fill[ck3] (3.5625,0.0000) rectangle (4.1250,1.5000);
\fill[pattern=north east lines, pattern color=white] (3.5625,0.0000) rectangle (4.1250,1.5000);
\draw[white,line width=0.6pt] (4.1250,0.0000) -- (4.1250,1.5000);
\fill[ck2] (4.1250,0.0000) rectangle (4.3125,1.5000);
\draw[white,line width=0.6pt] (4.3125,0.0000) -- (4.3125,1.5000);
\fill[ck0] (4.3125,0.0000) rectangle (4.5000,1.5000);
\fill[ck0] (4.5000,0.0000) rectangle (5.0625,1.5000);
\draw[white,line width=0.6pt] (5.0625,0.0000) -- (5.0625,1.5000);
\fill[ck2] (5.0625,0.0000) rectangle (5.6250,1.5000);
\draw[white,line width=0.6pt] (5.6250,0.0000) -- (5.6250,1.5000);
\fill[ck3] (5.6250,0.0000) rectangle (5.8125,1.5000);
\draw[white,line width=0.6pt] (5.8125,0.0000) -- (5.8125,1.5000);
\fill[ck1] (5.8125,0.0000) rectangle (6.0000,1.5000);
\draw[gray!70,thin] (0.0000,4.5000) rectangle (1.5000,6.0000);
\draw[gray!70,thin] (1.5000,4.5000) rectangle (3.0000,6.0000);
\draw[gray!70,thin] (3.0000,4.5000) rectangle (4.5000,6.0000);
\draw[gray!70,thin] (4.5000,4.5000) rectangle (6.0000,6.0000);
\draw[gray!70,thin] (0.0000,3.0000) rectangle (1.5000,4.5000);
\draw[gray!70,thin] (1.5000,3.0000) rectangle (3.0000,4.5000);
\draw[gray!70,thin] (3.0000,3.0000) rectangle (4.5000,4.5000);
\draw[gray!70,thin] (4.5000,3.0000) rectangle (6.0000,4.5000);
\draw[gray!70,thin] (0.0000,1.5000) rectangle (1.5000,3.0000);
\draw[gray!70,thin] (1.5000,1.5000) rectangle (3.0000,3.0000);
\draw[gray!70,thin] (3.0000,1.5000) rectangle (4.5000,3.0000);
\draw[gray!70,thin] (4.5000,1.5000) rectangle (6.0000,3.0000);
\draw[gray!70,thin] (0.0000,0.0000) rectangle (1.5000,1.5000);
\draw[gray!70,thin] (1.5000,0.0000) rectangle (3.0000,1.5000);
\draw[gray!70,thin] (3.0000,0.0000) rectangle (4.5000,1.5000);
\draw[gray!70,thin] (4.5000,0.0000) rectangle (6.0000,1.5000);
\draw[very thick] (0.0000,3.0000) rectangle (3.0000,6.0000);
\draw[very thick] (3.0000,3.0000) rectangle (6.0000,6.0000);
\draw[very thick] (0.0000,0.0000) rectangle (3.0000,3.0000);
\draw[very thick] (3.0000,0.0000) rectangle (6.0000,3.0000);
\end{tikzpicture}
\par\smallskip\textbf{(a)}
\end{minipage}\hfill
\begin{minipage}{0.325\textwidth}
\centering
\begin{tikzpicture}[x=0.8cm,y=0.8cm]
\fill[ck0] (0.0000,4.5000) rectangle (1.5000,6.0000);
\fill[ck1] (1.5000,4.5000) rectangle (3.0000,6.0000);
\fill[ck2] (3.0000,4.5000) rectangle (4.1250,6.0000);
\draw[white,line width=0.6pt] (4.1250,4.5000) -- (4.1250,6.0000);
\fill[ck3] (4.1250,4.5000) rectangle (4.5000,6.0000);
\fill[pattern=north east lines, pattern color=white] (4.1250,4.5000) rectangle (4.5000,6.0000);
\fill[ck3] (4.5000,4.5000) rectangle (5.6250,6.0000);
\fill[pattern=north east lines, pattern color=white] (4.5000,4.5000) rectangle (5.6250,6.0000);
\draw[white,line width=0.6pt] (5.6250,4.5000) -- (5.6250,6.0000);
\fill[ck2] (5.6250,4.5000) rectangle (6.0000,6.0000);
\fill[pattern=north east lines, pattern color=white] (5.6250,4.5000) rectangle (6.0000,6.0000);
\fill[ck2] (0.0000,3.0000) rectangle (1.5000,4.5000);
\fill[ck3] (1.5000,3.0000) rectangle (3.0000,4.5000);
\fill[ck0] (3.0000,3.0000) rectangle (4.1250,4.5000);
\fill[pattern=north east lines, pattern color=white] (3.0000,3.0000) rectangle (4.1250,4.5000);
\draw[white,line width=0.6pt] (4.1250,3.0000) -- (4.1250,4.5000);
\fill[ck1] (4.1250,3.0000) rectangle (4.5000,4.5000);
\fill[ck1] (4.5000,3.0000) rectangle (5.6250,4.5000);
\draw[white,line width=0.6pt] (5.6250,3.0000) -- (5.6250,4.5000);
\fill[ck0] (5.6250,3.0000) rectangle (6.0000,4.5000);
\fill[ck1] (0.0000,1.5000) rectangle (0.5625,3.0000);
\fill[pattern=north east lines, pattern color=white] (0.0000,1.5000) rectangle (0.5625,3.0000);
\draw[white,line width=0.6pt] (0.5625,1.5000) -- (0.5625,3.0000);
\fill[ck3] (0.5625,1.5000) rectangle (1.3125,3.0000);
\fill[pattern=north east lines, pattern color=white] (0.5625,1.5000) rectangle (1.3125,3.0000);
\draw[white,line width=0.6pt] (1.3125,1.5000) -- (1.3125,3.0000);
\fill[ck1] (1.3125,1.5000) rectangle (1.5000,3.0000);
\fill[pattern=north east lines, pattern color=white] (1.3125,1.5000) rectangle (1.5000,3.0000);
\fill[ck0] (1.5000,1.5000) rectangle (2.0625,3.0000);
\fill[pattern=north east lines, pattern color=white] (1.5000,1.5000) rectangle (2.0625,3.0000);
\draw[white,line width=0.6pt] (2.0625,1.5000) -- (2.0625,3.0000);
\fill[ck2] (2.0625,1.5000) rectangle (2.8125,3.0000);
\fill[pattern=north east lines, pattern color=white] (2.0625,1.5000) rectangle (2.8125,3.0000);
\draw[white,line width=0.6pt] (2.8125,1.5000) -- (2.8125,3.0000);
\fill[ck0] (2.8125,1.5000) rectangle (3.0000,3.0000);
\fill[pattern=north east lines, pattern color=white] (2.8125,1.5000) rectangle (3.0000,3.0000);
\fill[ck3] (3.0000,1.5000) rectangle (3.5625,3.0000);
\fill[pattern=north east lines, pattern color=white] (3.0000,1.5000) rectangle (3.5625,3.0000);
\draw[white,line width=0.6pt] (3.5625,1.5000) -- (3.5625,3.0000);
\fill[ck1] (3.5625,1.5000) rectangle (4.1250,3.0000);
\draw[white,line width=0.6pt] (4.1250,1.5000) -- (4.1250,3.0000);
\fill[ck0] (4.1250,1.5000) rectangle (4.3125,3.0000);
\draw[white,line width=0.6pt] (4.3125,1.5000) -- (4.3125,3.0000);
\fill[ck2] (4.3125,1.5000) rectangle (4.5000,3.0000);
\fill[pattern=north east lines, pattern color=white] (4.3125,1.5000) rectangle (4.5000,3.0000);
\fill[ck2] (4.5000,1.5000) rectangle (5.0625,3.0000);
\fill[pattern=north east lines, pattern color=white] (4.5000,1.5000) rectangle (5.0625,3.0000);
\draw[white,line width=0.6pt] (5.0625,1.5000) -- (5.0625,3.0000);
\fill[ck0] (5.0625,1.5000) rectangle (5.6250,3.0000);
\draw[white,line width=0.6pt] (5.6250,1.5000) -- (5.6250,3.0000);
\fill[ck1] (5.6250,1.5000) rectangle (5.8125,3.0000);
\fill[pattern=north east lines, pattern color=white] (5.6250,1.5000) rectangle (5.8125,3.0000);
\draw[white,line width=0.6pt] (5.8125,1.5000) -- (5.8125,3.0000);
\fill[ck3] (5.8125,1.5000) rectangle (6.0000,3.0000);
\fill[ck3] (0.0000,0.0000) rectangle (0.5625,1.5000);
\fill[pattern=north east lines, pattern color=white] (0.0000,0.0000) rectangle (0.5625,1.5000);
\draw[white,line width=0.6pt] (0.5625,0.0000) -- (0.5625,1.5000);
\fill[ck1] (0.5625,0.0000) rectangle (1.3125,1.5000);
\draw[white,line width=0.6pt] (1.3125,0.0000) -- (1.3125,1.5000);
\fill[ck3] (1.3125,0.0000) rectangle (1.5000,1.5000);
\fill[pattern=north east lines, pattern color=white] (1.3125,0.0000) rectangle (1.5000,1.5000);
\fill[ck2] (1.5000,0.0000) rectangle (2.0625,1.5000);
\fill[pattern=north east lines, pattern color=white] (1.5000,0.0000) rectangle (2.0625,1.5000);
\draw[white,line width=0.6pt] (2.0625,0.0000) -- (2.0625,1.5000);
\fill[ck0] (2.0625,0.0000) rectangle (2.8125,1.5000);
\draw[white,line width=0.6pt] (2.8125,0.0000) -- (2.8125,1.5000);
\fill[ck2] (2.8125,0.0000) rectangle (3.0000,1.5000);
\fill[pattern=north east lines, pattern color=white] (2.8125,0.0000) rectangle (3.0000,1.5000);
\fill[ck1] (3.0000,0.0000) rectangle (3.5625,1.5000);
\draw[white,line width=0.6pt] (3.5625,0.0000) -- (3.5625,1.5000);
\fill[ck3] (3.5625,0.0000) rectangle (4.1250,1.5000);
\draw[white,line width=0.6pt] (4.1250,0.0000) -- (4.1250,1.5000);
\fill[ck2] (4.1250,0.0000) rectangle (4.3125,1.5000);
\draw[white,line width=0.6pt] (4.3125,0.0000) -- (4.3125,1.5000);
\fill[ck0] (4.3125,0.0000) rectangle (4.5000,1.5000);
\fill[ck0] (4.5000,0.0000) rectangle (5.0625,1.5000);
\draw[white,line width=0.6pt] (5.0625,0.0000) -- (5.0625,1.5000);
\fill[ck2] (5.0625,0.0000) rectangle (5.6250,1.5000);
\draw[white,line width=0.6pt] (5.6250,0.0000) -- (5.6250,1.5000);
\fill[ck3] (5.6250,0.0000) rectangle (5.8125,1.5000);
\fill[pattern=north east lines, pattern color=white] (5.6250,0.0000) rectangle (5.8125,1.5000);
\draw[white,line width=0.6pt] (5.8125,0.0000) -- (5.8125,1.5000);
\fill[ck1] (5.8125,0.0000) rectangle (6.0000,1.5000);
\fill[pattern=north east lines, pattern color=white] (5.8125,0.0000) rectangle (6.0000,1.5000);
\draw[gray!70,thin] (0.0000,4.5000) rectangle (1.5000,6.0000);
\draw[gray!70,thin] (1.5000,4.5000) rectangle (3.0000,6.0000);
\draw[gray!70,thin] (3.0000,4.5000) rectangle (4.5000,6.0000);
\draw[gray!70,thin] (4.5000,4.5000) rectangle (6.0000,6.0000);
\draw[gray!70,thin] (0.0000,3.0000) rectangle (1.5000,4.5000);
\draw[gray!70,thin] (1.5000,3.0000) rectangle (3.0000,4.5000);
\draw[gray!70,thin] (3.0000,3.0000) rectangle (4.5000,4.5000);
\draw[gray!70,thin] (4.5000,3.0000) rectangle (6.0000,4.5000);
\draw[gray!70,thin] (0.0000,1.5000) rectangle (1.5000,3.0000);
\draw[gray!70,thin] (1.5000,1.5000) rectangle (3.0000,3.0000);
\draw[gray!70,thin] (3.0000,1.5000) rectangle (4.5000,3.0000);
\draw[gray!70,thin] (4.5000,1.5000) rectangle (6.0000,3.0000);
\draw[gray!70,thin] (0.0000,0.0000) rectangle (1.5000,1.5000);
\draw[gray!70,thin] (1.5000,0.0000) rectangle (3.0000,1.5000);
\draw[gray!70,thin] (3.0000,0.0000) rectangle (4.5000,1.5000);
\draw[gray!70,thin] (4.5000,0.0000) rectangle (6.0000,1.5000);
\draw[very thick] (0.0000,3.0000) rectangle (3.0000,6.0000);
\draw[very thick] (3.0000,3.0000) rectangle (6.0000,6.0000);
\draw[very thick] (0.0000,0.0000) rectangle (3.0000,3.0000);
\draw[very thick] (3.0000,0.0000) rectangle (6.0000,3.0000);
\end{tikzpicture}
\par\smallskip\textbf{(b)}
\end{minipage}\hfill
\begin{minipage}{0.325\textwidth}
\centering
\begin{tikzpicture}[x=0.8cm,y=0.8cm]
\fill[ck0] (0.0000,4.5000) rectangle (1.5000,6.0000);
\fill[ck1] (1.5000,4.5000) rectangle (3.0000,6.0000);
\fill[ck2] (3.0000,4.5000) rectangle (4.1250,6.0000);
\draw[white,line width=0.6pt] (4.1250,4.5000) -- (4.1250,6.0000);
\fill[ck3] (4.1250,4.5000) rectangle (4.5000,6.0000);
\fill[pattern=north east lines, pattern color=white] (4.1250,4.5000) rectangle (4.5000,6.0000);
\fill[ck3] (4.5000,4.5000) rectangle (5.6250,6.0000);
\fill[pattern=north east lines, pattern color=white] (4.5000,4.5000) rectangle (5.6250,6.0000);
\draw[white,line width=0.6pt] (5.6250,4.5000) -- (5.6250,6.0000);
\fill[ck2] (5.6250,4.5000) rectangle (6.0000,6.0000);
\fill[pattern=north east lines, pattern color=white] (5.6250,4.5000) rectangle (6.0000,6.0000);
\fill[ck2] (0.0000,3.0000) rectangle (1.5000,4.5000);
\fill[ck3] (1.5000,3.0000) rectangle (3.0000,4.5000);
\fill[ck0] (3.0000,3.0000) rectangle (4.1250,4.5000);
\draw[white,line width=0.6pt] (4.1250,3.0000) -- (4.1250,4.5000);
\fill[ck1] (4.1250,3.0000) rectangle (4.5000,4.5000);
\fill[pattern=north east lines, pattern color=white] (4.1250,3.0000) rectangle (4.5000,4.5000);
\fill[ck1] (4.5000,3.0000) rectangle (5.6250,4.5000);
\draw[white,line width=0.6pt] (5.6250,3.0000) -- (5.6250,4.5000);
\fill[ck0] (5.6250,3.0000) rectangle (6.0000,4.5000);
\fill[ck1] (0.0000,1.5000) rectangle (0.5625,3.0000);
\fill[pattern=north east lines, pattern color=white] (0.0000,1.5000) rectangle (0.5625,3.0000);
\draw[white,line width=0.6pt] (0.5625,1.5000) -- (0.5625,3.0000);
\fill[ck3] (0.5625,1.5000) rectangle (1.3125,3.0000);
\fill[pattern=north east lines, pattern color=white] (0.5625,1.5000) rectangle (1.3125,3.0000);
\draw[white,line width=0.6pt] (1.3125,1.5000) -- (1.3125,3.0000);
\fill[ck1] (1.3125,1.5000) rectangle (1.5000,3.0000);
\fill[pattern=north east lines, pattern color=white] (1.3125,1.5000) rectangle (1.5000,3.0000);
\fill[ck0] (1.5000,1.5000) rectangle (2.0625,3.0000);
\fill[pattern=north east lines, pattern color=white] (1.5000,1.5000) rectangle (2.0625,3.0000);
\draw[white,line width=0.6pt] (2.0625,1.5000) -- (2.0625,3.0000);
\fill[ck2] (2.0625,1.5000) rectangle (2.8125,3.0000);
\fill[pattern=north east lines, pattern color=white] (2.0625,1.5000) rectangle (2.8125,3.0000);
\draw[white,line width=0.6pt] (2.8125,1.5000) -- (2.8125,3.0000);
\fill[ck0] (2.8125,1.5000) rectangle (3.0000,3.0000);
\fill[pattern=north east lines, pattern color=white] (2.8125,1.5000) rectangle (3.0000,3.0000);
\fill[ck3] (3.0000,1.5000) rectangle (3.5625,3.0000);
\fill[pattern=north east lines, pattern color=white] (3.0000,1.5000) rectangle (3.5625,3.0000);
\draw[white,line width=0.6pt] (3.5625,1.5000) -- (3.5625,3.0000);
\fill[ck1] (3.5625,1.5000) rectangle (4.1250,3.0000);
\draw[white,line width=0.6pt] (4.1250,1.5000) -- (4.1250,3.0000);
\fill[ck0] (4.1250,1.5000) rectangle (4.3125,3.0000);
\draw[white,line width=0.6pt] (4.3125,1.5000) -- (4.3125,3.0000);
\fill[ck2] (4.3125,1.5000) rectangle (4.5000,3.0000);
\fill[pattern=north east lines, pattern color=white] (4.3125,1.5000) rectangle (4.5000,3.0000);
\fill[ck2] (4.5000,1.5000) rectangle (5.0625,3.0000);
\draw[white,line width=0.6pt] (5.0625,1.5000) -- (5.0625,3.0000);
\fill[ck0] (5.0625,1.5000) rectangle (5.6250,3.0000);
\fill[pattern=north east lines, pattern color=white] (5.0625,1.5000) rectangle (5.6250,3.0000);
\draw[white,line width=0.6pt] (5.6250,1.5000) -- (5.6250,3.0000);
\fill[ck1] (5.6250,1.5000) rectangle (5.8125,3.0000);
\draw[white,line width=0.6pt] (5.8125,1.5000) -- (5.8125,3.0000);
\fill[ck3] (5.8125,1.5000) rectangle (6.0000,3.0000);
\fill[pattern=north east lines, pattern color=white] (5.8125,1.5000) rectangle (6.0000,3.0000);
\fill[ck3] (0.0000,0.0000) rectangle (0.5625,1.5000);
\draw[white,line width=0.6pt] (0.5625,0.0000) -- (0.5625,1.5000);
\fill[ck1] (0.5625,0.0000) rectangle (1.3125,1.5000);
\fill[pattern=north east lines, pattern color=white] (0.5625,0.0000) rectangle (1.3125,1.5000);
\draw[white,line width=0.6pt] (1.3125,0.0000) -- (1.3125,1.5000);
\fill[ck3] (1.3125,0.0000) rectangle (1.5000,1.5000);
\fill[ck2] (1.5000,0.0000) rectangle (2.0625,1.5000);
\fill[pattern=north east lines, pattern color=white] (1.5000,0.0000) rectangle (2.0625,1.5000);
\draw[white,line width=0.6pt] (2.0625,0.0000) -- (2.0625,1.5000);
\fill[ck0] (2.0625,0.0000) rectangle (2.8125,1.5000);
\draw[white,line width=0.6pt] (2.8125,0.0000) -- (2.8125,1.5000);
\fill[ck2] (2.8125,0.0000) rectangle (3.0000,1.5000);
\fill[pattern=north east lines, pattern color=white] (2.8125,0.0000) rectangle (3.0000,1.5000);
\fill[ck1] (3.0000,0.0000) rectangle (3.5625,1.5000);
\draw[white,line width=0.6pt] (3.5625,0.0000) -- (3.5625,1.5000);
\fill[ck3] (3.5625,0.0000) rectangle (4.1250,1.5000);
\draw[white,line width=0.6pt] (4.1250,0.0000) -- (4.1250,1.5000);
\fill[ck2] (4.1250,0.0000) rectangle (4.3125,1.5000);
\draw[white,line width=0.6pt] (4.3125,0.0000) -- (4.3125,1.5000);
\fill[ck0] (4.3125,0.0000) rectangle (4.5000,1.5000);
\fill[ck0] (4.5000,0.0000) rectangle (5.0625,1.5000);
\draw[white,line width=0.6pt] (5.0625,0.0000) -- (5.0625,1.5000);
\fill[ck2] (5.0625,0.0000) rectangle (5.6250,1.5000);
\draw[white,line width=0.6pt] (5.6250,0.0000) -- (5.6250,1.5000);
\fill[ck3] (5.6250,0.0000) rectangle (5.8125,1.5000);
\fill[pattern=north east lines, pattern color=white] (5.6250,0.0000) rectangle (5.8125,1.5000);
\draw[white,line width=0.6pt] (5.8125,0.0000) -- (5.8125,1.5000);
\fill[ck1] (5.8125,0.0000) rectangle (6.0000,1.5000);
\fill[pattern=north east lines, pattern color=white] (5.8125,0.0000) rectangle (6.0000,1.5000);
\draw[gray!70,thin] (0.0000,4.5000) rectangle (1.5000,6.0000);
\draw[gray!70,thin] (1.5000,4.5000) rectangle (3.0000,6.0000);
\draw[gray!70,thin] (3.0000,4.5000) rectangle (4.5000,6.0000);
\draw[gray!70,thin] (4.5000,4.5000) rectangle (6.0000,6.0000);
\draw[gray!70,thin] (0.0000,3.0000) rectangle (1.5000,4.5000);
\draw[gray!70,thin] (1.5000,3.0000) rectangle (3.0000,4.5000);
\draw[gray!70,thin] (3.0000,3.0000) rectangle (4.5000,4.5000);
\draw[gray!70,thin] (4.5000,3.0000) rectangle (6.0000,4.5000);
\draw[gray!70,thin] (0.0000,1.5000) rectangle (1.5000,3.0000);
\draw[gray!70,thin] (1.5000,1.5000) rectangle (3.0000,3.0000);
\draw[gray!70,thin] (3.0000,1.5000) rectangle (4.5000,3.0000);
\draw[gray!70,thin] (4.5000,1.5000) rectangle (6.0000,3.0000);
\draw[gray!70,thin] (0.0000,0.0000) rectangle (1.5000,1.5000);
\draw[gray!70,thin] (1.5000,0.0000) rectangle (3.0000,1.5000);
\draw[gray!70,thin] (3.0000,0.0000) rectangle (4.5000,1.5000);
\draw[gray!70,thin] (4.5000,0.0000) rectangle (6.0000,1.5000);
\draw[very thick] (0.0000,3.0000) rectangle (3.0000,6.0000);
\draw[very thick] (3.0000,3.0000) rectangle (6.0000,6.0000);
\draw[very thick] (0.0000,0.0000) rectangle (3.0000,3.0000);
\draw[very thick] (3.0000,0.0000) rectangle (6.0000,3.0000);
\end{tikzpicture}
\par\smallskip\textbf{(c)}
\end{minipage}
\caption{Signed fractional Sudoku over the unsigned pattern of
Figure~\ref{fig:ex-pair}(b); a field is hatched when the corresponding entry is
negative. (a)~$\mathbf P=T(\pi/4,\pi/3)$. (b)~$\mathbf P'$, obtained with the left
factor $\Gamma$ only. (c)~$\mathbf P''$, with in addition the right factors
$D_U,D_V,D_W$: the same SudoQ as $\mathbf P'$, with the vectors in the cells
$(2,3)$, $(3,4)$ and $(4,1)$ reversed.}
\label{fig:app-signed}
\end{figure}
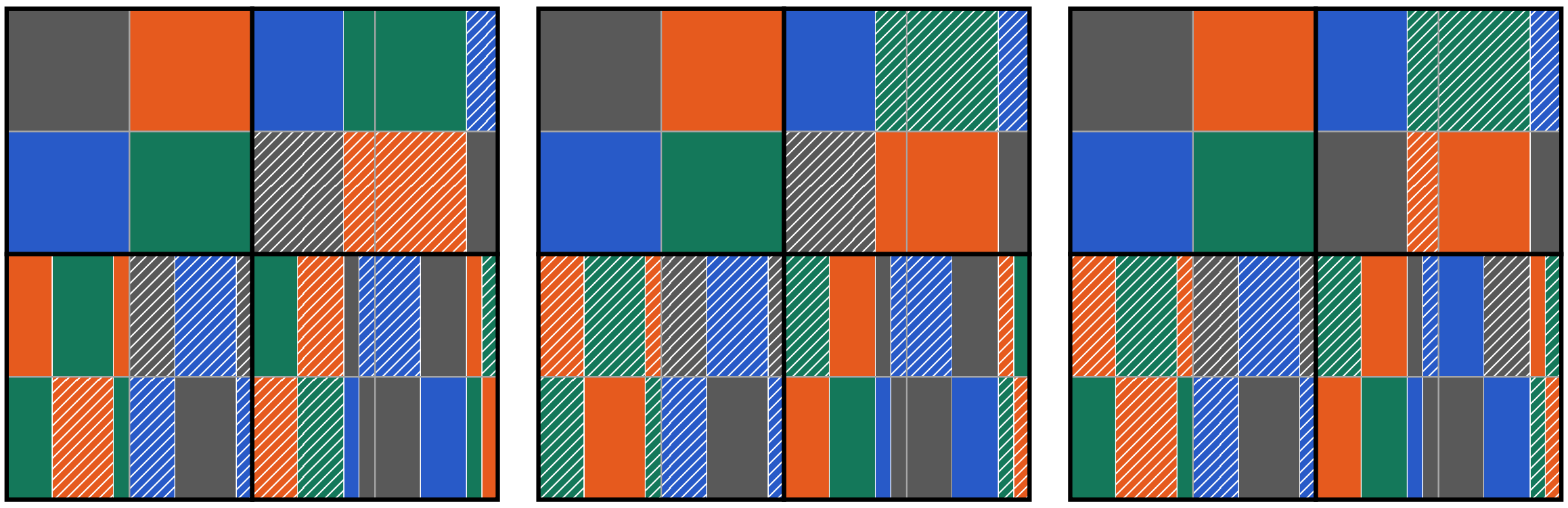

The two kinds of factors act differently on the picture, as shown in
Figure~\ref{fig:app-signed}. The left factor $\Gamma$ changes the signs of rows $2$ and $4$
of $U,V,W$: from (a) to (b), the hatching of the orange and
green fields is reversed in the three corresponding blocks.
In the first block, the canonical projections are unchanged,
and we keep their positive vector representatives
$e_1,e_2,e_3,e_4$, so this block remains unhatched.
A right factor changes the sign of a column: from (b) to (c),
the hatching of every coloured region is reversed in the cells
$(2,3)$, $(4,1)$ and $(3,4)$, without changing the corresponding
projections.

All three panels have the same unsigned pattern, as they must: by
Theorem~\ref{thm:fs}(1) it determines the class of the grid. The hatching does
not. It depends on the vectors chosen to represent the projections, and on the
basis. By Proposition 5.2, every real normalised grid equivalent to
$\mathbf P$ is obtained by conjugating all its projections
by a diagonal unitary $D=\operatorname{diag}(d_1,d_2,d_3,d_4)$.
In this example, $W$ is real and all its entries are nonzero.
Thus, for any column $w$ of $W$, the reality of $Dww^*D^*$
forces $d_i\overline{d_j}$ to be real for all
$i,j\in\{1,2,3,4\}$.
Up to an irrelevant common phase, $D$ is therefore a diagonal
sign matrix $\Gamma=\operatorname{diag}(\pm1,\pm1,\pm1,\pm1)$.
Moreover, since every component of $w$ is nonzero,
$\Gamma ww^*\Gamma=ww^*$ implies $\Gamma=\pm I_4$.
Consequently, the $2^4$ diagonal sign matrices give exactly
$2^4/2=8$ distinct real normalised grids in the class of
$\mathbf P$, one of which is $\mathbf P'=\mathbf P''$.

\FloatBarrier

\section*{Statements and Declarations}

\subsection*{Funding}
{\sloppy
This research is supported by Gruppo Nazionale di Fisica Matematica
GNFM-INdAM, Istituto Nazionale di Fisica Nucleare INFN through the project
QUANTUM, and financially supported by 
%PNRR MUR project PE0000023-NQSTI,
PNRR MUR project CN00000013 `Italian National Centre on HPC, Big Data and
Quantum Computing', PRIN 2022 project 2022TEB52W-PE1- `The charm of
integrability: from nonlinear waves to random matrices', and MUR project
PRIN 2022SC, Grant ref.~2022RFAZCJ. 
FDC,  PF, ML acknowledge support from Projects PN-RIC Q-SUD (Q-SUD B99H26000380007) and PN-RIC QUANTAS (SYNERGIA B99H26000410007).
K{\.Z}~acknowledges financial support
from the European Union through ERC Advanced Grant \emph{TAtypic}, Project
No.~101142236.
\par}

\subsection*{Competing interests}
The authors have no relevant financial or non-financial interests to disclose.

\subsection*{AI use disclosure}
AI assistance was used for editing, condensing the
exposition and the presentation of the proofs, and for preparing the
figures. All results and arguments were checked by the authors.


\begin{thebibliography}{99}
\bibitem{AFLT}
E. Arnold, R. Field, S. Lucas, and L. Taalman,
\emph{Minimal complete Shidoku symmetry groups},
J. Combin. Math. Combin. Comput. \textbf{87}, 209--228 (2013).

\bibitem{BvN}
G. Birkhoff and J. von Neumann,
\emph{The Logic of Quantum Mechanics},
Annals of Mathematics \textbf{37}, 823--843 (1936).

\bibitem{Bistron26}
R. Bistro\'n and J. Czartowski,
\emph{Local thermal operations and classical communication},
Phys. Rev. E \textbf{113}, 034114 (2026).

\bibitem{DHW2000}
J. K. Dugdale, A. J. W. Hilton, and J. Wojciechowski,
\emph{Fractional Latin squares, simplex algebras, and generalized quotients},
J. Statist. Plann. Inference \textbf{86}, 457--504 (2000).

\bibitem{DN2024}
P. J. Dukes and K. I. Nimegeers,
\emph{The linear system for Sudoku and a fractional completion threshold},
Algebr. Comb. \textbf{7}, 1283--1305 (2024).

\bibitem{Felgenhauer06}
B. Felgenhauer and A. F. Jarvis,
\emph{Mathematics of Sudoku I},
Math. Spectrum \textbf{39}, 15--22 (2006).

\bibitem{HZ2026}
M. Hontarenko and K. \.Zyczkowski,
\emph{Fractional and quantum Latin squares: real solutions of quantum
36 officers problem}, arXiv:2610.17170 (2026).

\bibitem{McGuire14}
G. McGuire, B. Tugemann, and G. Civario,
\emph{There is no 16-clue Sudoku: solving the Sudoku minimum number of
clues problem via hitting set enumeration},
Exp. Math. \textbf{23}, 190--217 (2014).

\bibitem{Munkres}
J. R. Munkres, \emph{Topology}, 2nd ed., Prentice Hall, 2000.

\bibitem{MustoMUB}
B. Musto,
\emph{Constructing mutually unbiased bases from quantum Latin squares},
Electron. Proc. Theor. Comput. Sci. \textbf{236}, 108--126 (2017).

\bibitem{MV}
B. Musto and J. Vicary,
\emph{Quantum Latin squares and unitary error bases},
Quantum Inf. Comput. \textbf{16}, 1318--1332 (2016).

\bibitem{MVOrth}
B. Musto and J. Vicary,
\emph{Orthogonality for quantum Latin isometry squares},
Electron. Proc. Theor. Comput. Sci. \textbf{287}, 253--266 (2019).

\bibitem{NP}
I. Nechita and J. Pillet,
\emph{SudoQ---a quantum variant of the popular game},
Quantum Inf. Comput. \textbf{21}, 781--799 (2021).

\bibitem{KZ}
J. Paczos, M. Wierzbi\'nski, G. Rajchel-Mieldzio\'c,
A. Burchardt, and K. \.Zyczkowski,
\emph{Genuinely quantum solutions of the game Sudoku and their cardinality},
Phys. Rev. A \textbf{104}, 042423 (2021).

\bibitem{RBB2022}
S. A. Rather, A. Burchardt, W. Bruzda, G. Rajchel-Mieldzio\'c,
A. Lakshminarayan, and K. \.Zyczkowski,
\emph{Thirty-six entangled officers of Euler: quantum solution to a
classically impossible problem},
Phys. Rev. Lett. \textbf{128}, 080507 (2022).

\bibitem{Russell06}
E. Russell and A. F. Jarvis,
\emph{Mathematics of Sudoku II},
Math. Spectrum \textbf{39}, 54--58 (2006).

\bibitem{Zang_2021}
Y. Zang, P. Facchi, and Z. Tian
\emph{Quantum combinatorial designs and $k$-uniform states},
J. Phys. A: Math. Theor. \textbf{54}, 505204 1--33 (2021).


\bibitem{ZZTF}
Y. Zang, H. Zhang, Z. Tian, and S.-M. Fei,
\emph{Genuinely quantum solutions of orthogonal Sudoku squares},
Adv. Quantum Technol. \textbf{9}, e70293 (2026).

\end{thebibliography}
\end{document}